# On the accuracy of the GN-model and on analytical correction terms to improve it

Andrea Carena,[1] Gabriella Bosco,[1] Vittorio Curri,[1]
Yanchao Jiang,[1] Pierluigi Poggiolini,[1,*] and Fabrizio Forghieri[2]

[1]*Dipartimento di Elettronica e Telecomunicazioni, Politecnico di Torino,
Corso Duca degli Abruzzi 24, 10129, Torino, Italy*
[2]*CISCO Photonics, Via Philips 12, 20052, Monza, Milano, Italy,*
*[*pierluigi.poggiolini@polito.it](mailto:pierluigi.poggiolini@polito.it)

**Abstract:** The GN-model has been proposed as an approximate but sufficiently accurate tool for predicting uncompensated optical coherent transmission system performance, in realistic scenarios. For this specific use, the GN-model has enjoyed substantial validation, both simulative and experimental. Recently, however, it has been pointed out that its predictions, when used to obtain a detailed picture of non-linear interference (NLI) noise accumulation along a link, may be affected by a substantial NLI overestimation error, especially in the first spans of the link. In this paper we analyze in detail the GN-model errors. We discuss recently proposed formulas for correcting such errors and show that they neglect several contributions to NLI, so that they may substantially underestimate NLI in specific situations, especially over low-dispersion fibers. We derive a complete set of formulas accounting for all single, cross, and multi-channel effects, this set constitutes what we have called the enhanced GN-model (EGN-model). We extensively validate the EGN model by comparison with accurate simulations in several different system scenarios. The overall EGN model accuracy is found to be very good when assessing detailed span-by-span NLI accumulation and excellent when estimating realistic system maximum reach. The computational complexity vs. accuracy trade-offs of the various versions of the GN and EGN models are extensively discussed.

## References and links

## 1. Introduction

Building on results from several similar prior modeling efforts [1-5], the GN-model of non-linear fiber propagation has recently been proposed [6-14]. A more extensive bibliography and a comprehensive model description are provided in [11,14].

Since the start, the GN-model main purpose has declaredly been that of providing a simple but sufficiently accurate tool for the prediction of the main system performance indicators in uncompensated links that make use of coherent detection. Typical such indicators are maximum reach and optimum launch power. For this specific use, the GN-model has obtained substantial validation, both simulative [6,7,14,15,21] and experimental [16-21], by various independent groups.

Recently, however, it has been pointed out that when the GN-model is used to look at the detailed span-by-span characterization of non-linear interference (NLI) accumulation along a link, its predictions may be affected by a substantial error [22-25]. In particular in [22], the first peer-reviewed published paper on the subject (simultaneously with [23]), we presented for the first time a detailed picture of the predicted and actual NLI noise variance accumulated along realistic links based on polarization-multiplexed (PM) QPSK and PM-16QAM. We showed that the GN-model overestimates the variance of NLI, most notably in the first spans of the link, where this error may amount to several dB's, depending on system parameters and modulation format. The error then abates considerably along the link, but it does not vanish. We showed this error to be related to one of the GN-model main approximations: the 'signal Gaussianity' assumption, which consists in assuming that the transmitted signal, due to uncompensated dispersion, approximately behaves as Gaussian noise. Especially in the first spans of the link, this approximation is not accurate and generates substantial error.

Independently of [22], another paper [24] later focused on the issue of the GN-model accuracy. Remarkably, [24] succeeded in analytically removing the signal Gaussianity assumption. A 'correction term' to the GN-model, limited to XPM (cross-phase modulation), was found. The results of [24] constitute major progress, also because it was shown that removing the signal Gaussianity assumption does not lead to unmanageably complex calculations, as we previously believed.

In this paper we adopt a similar approach to that indicated in [24] and in Sect. 3 we provide for the first time the GN-model 'correction terms' for single-channel non linearity (that is, self-channel interference or SCI), which was not addressed in [24]. In Sect. 4 we provide the formulas for the NLI noise due to XCI (cross-channel interference) and show them to contain more contributions than accounted for in the XPM formulas of [24]. In Sect. 5 we discuss the impact of MCI (multi-channel interference), which was neglected in [24], and show it to contribute substantially to NLI in certain specific scenarios, namely with low-dispersion fibers such as TrueWave RS or LS. We provide the formulas needed to account for MCI as well. Overall, we supply a complete set of equations that fully correct the GN-model for the effect of signal non-Gaussianity. We call this overall set of equations *the enhanced GN-model* (EGN-model). We carefully compare the EGN-model predictions with accurate simulations of span-by-span NLI accumulation and find the EGN-model accuracy to be very good. We also find that the XPM formulas proposed in [24] may in certain cases substantially underestimate NLI, especially with low-dispersion fibers. This circumstance is extensively discussed in both Sect. 4 and 5.

In Sect. 6 we apply the EGN-model to various realistic system scenarios involving PM-QPSK and PM-16QAM. Specifically, we concentrate on a comparison of the estimate of system maximum reach obtained using either the GN-model or the EGN-model, vs. accurate simulation results. Our bottom-line findings are that, when used for predicting realistic PM-QAM systems maximum reach at 32 GBaud, the GN-model error is always conservative, i.e., it underestimates the maximum reach, by typically 0.3-0.6 dB, and up to 0.8 dB over ultra-low dispersion fibers such as Corning LS. The EGN-model provides much better accuracy, completely removing the underestimation incurred by the GN-model. The error range across all considered fibers and channel spacing values is reduced to less than 0.2 dB. Such error range is so low that it becomes difficult to attribute it to either residual model inaccuracy or Monte-Carlo simulation uncertainty.

The resulting complexity of the EGN-model is however rather large and in Sect. 7 we discuss the issue of computational effort for realistic system performance prediction, providing a set of guidelines and recommendations. We point out that the very simple 'incoherent' GN-model [6,7,14] possibly represents an attractive compromise between accuracy and complexity, providing rather precise maximum reach predictions in many practical scenarios with small computational effort. We also point out that, for the purpose of system performance studies, an analytical closed-form GN-model correction formula, based on an approximation of the EGN-model, has been proposed in [27]. This approximation adds little complexity to that of the GN-model and substantially improves its accuracy. However, if ultra-accurate system performance prediction is critical or when a span-by-span detailed picture of NLI is of interest, then the full EGN-model presented here must be used.

In Sect. 7B we discuss the presence and relevance of *phase noise* within NLI, an aspect that was addressed in [24,25]. Our results indicate that, in the context of realistic systems, phase noise appears to have small or negligible impact on system performance prediction. In other words, the assumption of NLI noise being Gaussian and additive appears to be adequate for system performance predictions in most practical system scenarios. In Sect. 7C we briefly address the issue of modeling results vs. experimental evidence. Comments and conclusion follow.

In the following, for simplicity we call 'GN-model' the coherent-NLI-accumulation GN-model described in [14] as Eq. (1). We call 'incoherent GN-model' the simplified GN-model version that assumes incoherent NLI accumulation, described in [14] as Eq. (9).

## 2.  The EGN-model

When removing the assumption that the signal launched into the link statistically behaves as Gaussian noise, the power spectral density (PSD) of NLI turns out to be expressed by two terms:

$$G_{\mathrm{NLI}}^{\mathrm{EGN}}(f) = G_{\mathrm{NLI}}^{\mathrm{GN}}(f) + G_{\mathrm{NLI}}^{\mathrm{corr}}(f) \tag{1}$$

The first term, $G_{\mathrm{NLI}}^{\mathrm{GN}}(f)$, is the GN-model. The other, $G_{\mathrm{NLI}}^{\mathrm{corr}}(f)$, can be thought of as a *correction term* which takes the effect of signal non-Gaussianity into account. In the following, we call the overall resulting corrected model $G_{\mathrm{NLI}}^{\mathrm{EGN}}(f)$ as the 'enhanced GN-model', or EGN-model.

In this section we orderly present the EGN-model formulas according to the type of NLI that they address, namely SCI, XCI and MCI. In other words, we will break down $G_{\mathrm{NLI}}^{\mathrm{EGN}}(f)$ as:

$$G_{\mathrm{NLI}}^{\mathrm{EGN}}(f) = G_{\mathrm{SCI}}^{\mathrm{EGN}}(f) + G_{\mathrm{XCI}}^{\mathrm{EGN}}(f) + G_{\mathrm{MCI}}^{\mathrm{EGN}}(f) \tag{2}$$

Note that each one of the right-hand side terms possesses both a GN-model part and a correction part, in agreement with Eq. (1). For instance: $G_{\mathrm{SCI}}(f) = G_{\mathrm{SCI}}^{\mathrm{GN}}(f) + G_{\mathrm{SCI}}^{\mathrm{corr}}(f)$, and similarly for $G_{\mathrm{XCI}}^{\mathrm{EGN}}(f)$ and $G_{\mathrm{MCI}}^{\mathrm{EGN}}(f)$. We will point out which is which in their defining formulas.

The reason for resorting in Eq. (2) to this subdivision of NLI contributions is that it more naturally relates to the GN-model than the traditional taxonomy. Before proceeding, we recall the definition of the three NLI types, for the readers' convenience. An equivalent but more formal set of definitions, based on the actual spectral position of the WDM signal components beating together, can be found in [11], Sect. VI. The NLI impinging on the channel-under-test (CUT) of a WDM comb is the sum of three types of NLI contributions:

- Self-channel interference (SCI): it is NLI caused by the CUT on itself.
- Cross-channel interference (XCI): it is NLI affecting the CUT caused by the beating of the CUT with any single interfering (INT) channel.
- Multiple-channel interference (MCI): it is NLI affecting the CUT, caused by the beating of the CUT with two INT channels simultaneously, or the beating of three INT channels simultaneously.

In the following, we assume a multi-span link, with lumped amplification and all identical spans. We assume *dual polarization throughout.* We also assume that channels have rectangular spectra with same bandwidth, equal to the symbol rate $R_s$. These limiting assumptions could be removed but they are kept here to avoid excessive complexity in the resulting formulas. Specifically, note that if the transmitted channel spectra are not rectangular, the integer parameter $p$ introduced in Appendix E, can be non-zero and more GN-model correction terms are generated. We do not address this case in this paper.

The main symbols used in this paper are defined in the following, with units that make the subsequent formulas self-consistent:

- $z$ : the longitudinal spatial coordinate, along the link [km]
- $\alpha$ : optical field fiber loss [1/km], such that the optical *field* attenuates as $e^{-\alpha z}$; note that the optical *power* attenuates as $e^{-2\alpha z}$
- $\beta_2$ : dispersion coefficient [ps$^2$/km]
- $\gamma$ : fiber non-linearity coefficient [1/(W km)]
- $L_s$ : span length [km]
- $L_{\text{eff}} = \left(1 - e^{-2\alpha L_s}\right)/2\alpha$ : span effective length [km]
- $N_s$ : total number of spans in a link
- $R_s$ : symbol rate of an individual channel [TBaud]
- $T_s$ : duration of a symbol, equal to $R_s^{-1}$ [ps]
- $s_{\text{CUT}}(t)$ , $s_{\text{INT}}(t)$ : the pulses used by either the CUT or the INT channels, respectively
- $s_{\text{CUT}}(f)$ , $s_{\text{INT}}(f)$ : Fourier transforms of the above, assumed rectangular with bandwidth $R_s$ and flat-top value equal to $1/R_s$, centered at $f = 0$ for the CUT and at $f = f_c$ for an INT channel. Note that for these signals we use the same symbol both in frequency-domain and in time-domain. They can be distinguished based on the argument which is either a time $t$ or a frequency $f$.
- $f_c$ : the center frequency of an INT channel, such that the CUT and INT channel do not overlap [THz]
- $a_x$ , $a_y$ , $b_x$ , $b_y$ : random variables (RVs), representing the generic symbols transmitted in either the CUT ('$a$' RVs), or the INT channels ('$b$' RVs), respectively, on either polarization (subscripts $x$ and $y$)

According to the above definitions, the CUT overall transmitted signal can be written as:

$$S_{\text{CUT}}(t) = \sum_n \left( a_{x,n}\hat{x} + a_{y,n}\hat{y} \right) s_{\text{CUT}}\left( t - nT_s \right) \tag{3}$$

and similarly for the INT channel, with '$b$' RV's in the formula. The average transmitted power in the CUT and INT channels are given by:

$$P_{\text{CUT}} = \mathrm{E}\left\{ \left| a_x \right|^2 + \left| a_y \right|^2 \right\} \quad , \quad P_{\text{INT}} = \mathrm{E}\left\{ \left| b_x \right|^2 + \left| b_y \right|^2 \right\} \tag{4}$$

### 3. Self-channel interference (SCI)

The NLI produced by a CUT onto itself is SCI. Its contribution can be rather substantial. In a densely packed, full C-band system, operating at 32 GBaud, it approximately ranges between 20% and 40% of the total NLI power perturbing the CUT, over a wide range of fiber parameters and link lengths.

In [24] SCI was not dealt with and all calculations/simulations assumed that SCI was removed. In theory, removing SCI may be possible using electronic non-linear-compensation (NLC). While NLC is a fervid field of investigation, so far it is unclear whether NLC can be effectively implemented in DSP. At present, there are no commercial products incorporating it. Therefore, it seems quite necessary to include SCI as well, in dealing with a GN-model upgrade.

To derive the SCI formulas we used an approach similar to [24], suitably taking into account the effect of the non-Gaussianity of the signal. The derivation is shown in Appendix A. The NLI power spectral density (PSD) emerging at a generic frequency $f$ within the CUT, due to the interference of a single channel onto itself, in dual polarization, is given by:

$$G_{\text{SCI}}^{\text{EGN}}(f) = P_{\text{SCI}}^3 \left[ \kappa_1(f) + \Phi_a \, \kappa_2(f) + \Psi_a \, \kappa_3(f) \right] \tag{5}$$

where:

$$\Phi_a = \frac{\mathrm{E}\left\{ \left| a \right|^4 \right\}}{\mathrm{E}^2\left\{ \left| a \right|^2 \right\}} - 2 \quad , \quad \Psi_a = \frac{\mathrm{E}\left\{ \left| a \right|^6 \right\}}{\mathrm{E}^3\left\{ \left| a \right|^2 \right\}} - 9\frac{\mathrm{E}\left\{ \left| a \right|^4 \right\}}{\mathrm{E}^2\left\{ \left| a \right|^2 \right\}} + 12 \tag{6}$$

$$\kappa_1(f) = \frac{16}{27} R_s^3 \int_{-R_s/2}^{+R_s/2} df_1 \int_{-R_s/2}^{+R_s/2} df_2 \left| s_{\text{CUT}}(f_1) \right|^2 \left| s_{\text{CUT}}(f_2) \right|^2 \left| s_{\text{CUT}}(f_1 + f_2 - f) \right|^2 \left| \mu(f_1, f_2, f) \right|^2 \tag{7}$$

$$
\begin{aligned}
\kappa_2(f) = {} & \frac{80}{81} R_s^2 \int_{-R_s/2}^{+R_s/2} df_1 \int_{-R_s/2}^{+R_s/2} df_2 \int_{-R_s/2}^{+R_s/2} df_2' \\
& \left| s_{\text{CUT}}(f_1) \right|^2 s_{\text{CUT}}(f_2) s_{\text{CUT}}^*(f_2') s_{\text{CUT}}^*(f_1 + f_2 - f) s_{\text{CUT}}(f_1 + f_2' - f) \mu(f_1, f_2, f) \mu^*(f_1, f_2', f) \\
& + \frac{16}{81} R_s^2 \int_{-R_s/2}^{+R_s/2} df_1 \int_{-R_s/2}^{+R_s/2} df_2 \int_{-R_s/2}^{+R_s/2} df_2' \\
& \left| s_{\text{CUT}}(f_1 + f_2 - f) \right|^2 s_{\text{CUT}}(f_1) s_{\text{CUT}}(f_2) s_{\text{CUT}}^*(f_1 + f_2 - f_2') s_{\text{CUT}}^*(f_2') \mu(f_1, f_2, f) \mu^*(f_1 + f_2 - f_2', f_2', f)
\end{aligned}
\tag{8}
$$

$$
\begin{aligned}
\kappa_3(f) = {} & \frac{16}{81} R_s \int_{-R_s/2}^{+R_s/2} df_1 \int_{-R_s/2}^{+R_s/2} df_2 \int_{-R_s/2}^{+R_s/2} df_1' \int_{-R_s/2}^{+R_s/2} df_2' \, s_{\text{CUT}}(f_1) s_{\text{CUT}}(f_2) \\
& s_{\text{CUT}}^*(f_1 + f_2 - f) s_{\text{CUT}}^*(f_1') s_{\text{CUT}}^*(f_2') s_{\text{CUT}}(f_1' + f_2' - f) \, \mu(f_1, f_2, f) \mu^*(f_1', f_2', f)
\end{aligned}
\tag{9}
$$

where $\mu(f_1, f_2, f)$ is a 'link function' which weighs the generation of NLI and depends only on the link parameters, but not on the characteristics of the launched signal.

Under the assumptions made in this paper of lumped amplification and identical spans, the factor $\mu$ can be written as:

$$\mu(f_1, f_2, f) = \zeta(f_1, f_2, f) \cdot \nu(f_1, f_2, f) \tag{10}$$

where:

$$\zeta(f_1, f_2, f) = \gamma \frac{1 - e^{-2\alpha L_s} e^{j4\pi^2 \beta_2 (f_1 - f)(f_2 - f) L_s}}{2\alpha - j4\pi^2 \beta_2 (f_1 - f)(f_2 - f)} \tag{11}$$

$$\nu(f_1, f_2, f) = \frac{\sin\left(2\beta_2 \pi^2 (f_1 - f)(f_2 - f) N_s L_s\right)}{\sin\left(2\beta_2 \pi^2 (f_1 - f)(f_2 - f) L_s\right)} e^{j2\beta_2 \pi^2 (f_1 - f)(f_2 - f)(N_s - 1) L_s} \tag{12}$$

The $\zeta(f_1, f_2, f)$ factor physically represents the efficiency of non-degenerate four-wave mixing (FWM) occurring among three spectral components of the signal placed at frequencies $f_1$, $f_2$, $f_3 = (f_1 + f_2 - f)$, producing a beat disturbance at frequency $f$. The factor $\nu$ relates to the coherent interference of the NLI field contributions produced in different spans, when they are summed up at the receiver location. For more details on these factors, see [10,14] and the appendices of this paper.

If distributed amplification or non-identical spans are present in the link, the EGN formulas shown in this paper are still valid, provided that the link function $\mu$ is suitably modified. These generalizations will not be dealt with in this paper. However, for the interested readers, [10] Eq. (100) provides $\mu(f_1, f_2, f)$ for arbitrarily different lumped-amplification spans, whereas [10], p. 16, Eq. (I.2), further generalizes it to arbitrarily different distributed-amplification spans. In both cases, $\mu(f_1, f_2, f)$ is the expression contained within absolute value squared. To use it here, the absolute value squared must be removed.

The term related to $\kappa_1(f)$ in Eq. (5) accounts for the GN-model component, that is: $G_{\text{SCI}}^{\text{GN}}(f) = P_{\text{SCI}}^3 \kappa_1(f)$. The other two terms are corrections that take signal non-Gaussianity into account, that is: $G_{\text{SCI}}^{\text{corr}}(f) = P_{\text{SCI}}^3 \left[\Phi_a \kappa_2(f) + \Psi_a \kappa_3(f)\right]$. Note the need to include both a $4^{\text{th}}$ and a $6^{\text{th}}$-order moment of the transmitted symbol sequence, the latter appearing in the coefficient $\Psi_a$. The values of $\Phi_a$ and $\Psi_a$ depend only on the chosen format. In Table 1 we report them for the most common QAM constellations. It shows that more complex formats have smaller values of $\Phi_a$ and $\Psi_a$. As a result, they have a smaller correction $G_{\text{SCI}}^{\text{corr}}(f)$ vs. the GN-model component $G_{\text{SCI}}^{\text{GN}}(f)$. This is also the case with XCI and MCI (Sects. 4 and 5).

**Table 1: values of $\Phi_a$ and $\Psi_a$**

| Format | $\Phi_a$ | $\Psi_a$ |
|--------|--------|--------|
| BPSK | -1 | 4 |
| QPSK | -1 | 4 |
| 16QAM | -17/25 | 52/25 |
| 64 QAM | -13/21 | 1161/646 |

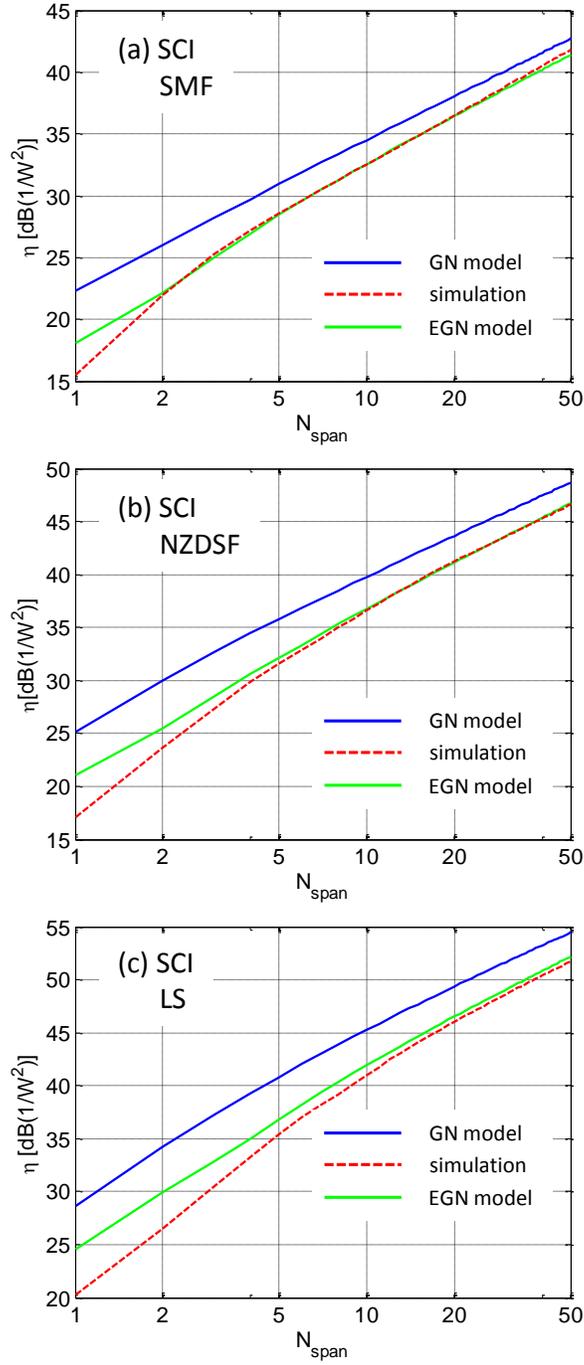

Fig. 1: Plot of normalized Self-Channel Interference (SCI), $\eta_{SCI}$, vs. number of spans in the link, assuming a single PM-QPSK channel over (from top to bottom) SMF, NZDSF and LS, with span length 100 km. Red dashed line: simulation. Blue solid line: GN-model. Green solid line: EGN-model (Eq. (5)).

In Figs. 1(a)-1(c) we show the result of the SCI calculation vs. simulations. Details about the simulation technique can be found in [14], where similar simulations were carried out. The simulated data length amounted to 300,000 symbols, a number that was used for all NLI span-by-span accumulation plots in this paper. We looked at the SCI normalized average power $\eta_{\text{SCI}}$ defined as follows:

$$\eta_{\text{SCI}} = P_{\text{CUT}}^{-3} \int_{-R_s/2}^{R_s/2} G_{\text{SCI}}^{\text{EGN}}(f)\,df \qquad (13)$$

This parameter collects the total SCI noise spectrally located over the CUT, normalized through $P_{\text{CUT}}^{-3}$ so that $\eta_{\text{SCI}}$ itself does not depend on launch power. The simulated system data are as follows:

- single channel PM-QPSK at $R_s$ =32 GBaud
- raised-cosine power spectrum with roll-off parameter 0.05
- SMF with $D$ =16.7 [ps/(nm km)], $\gamma$ =1.3 [1/(W km)], $\alpha_{\text{dB}}$ =0.22 [dB/km]
- NZDSF with $D$ =3.8 [ps/(nm km)], $\gamma$ =1.5 [1/(W km)], $\alpha_{\text{dB}}$ =0.22 [dB/km]
- LS fiber with $D$ =-1.8 [ps/(nm km)], $\gamma$ =2.2 [1/(W km)], $\alpha_{\text{dB}}$ =0.22 [dB/km]
- span length $L_s$ =100 [km]

Note that we chose not to use ideally rectangular spectra, to avoid possible numerical problems due to the truncation of excessively long, slowly decaying signal pulses. The selected roll-off value is very small and non-linearity generation can be expected not to differ significantly from that of an ideal rectangular spectrum. We chose PM-QPSK as modulation format to maximize the correction $G_{\text{SCI}}^{\text{corr}}(f)$ vs. the GN-model term $G_{\text{SCI}}^{\text{GN}}(f)$, according to Table 1. The same format was used, for the same reason, for the investigation of XCI and MCI span-by-span accumulation, shown in Sects. 4 and 5.

The plots in Fig. 1 show that Eq. (5) has good accuracy, as soon as there is some substantial accumulated dispersion. The gap between analytical and simulative results in the first few spans is currently being investigated. Beyond the first few spans, the agreement is excellent for SMF and NZDSF and still rather good for the challenging, very low-dispersion LS fiber. The overall accuracy improvement over the GN-model is very substantial.

Note also that the difference between either simulation or the EGN-model, vs. the GN-model (blue line) tends to close up for large number of spans. At 50 spans the residual gap is 1.1 dB for SMF. It is however more significant for the lower-dispersion fibers: 2.1 dB for NZDSF and 2.8 dB for LS.

## 4. Cross-channel interference (XCI)

A key aspect of XCI is that the individual contributions of each single INT channel in the WDM comb simply add up. As a result, one can concentrate on analytically finding the XCI due to a single INT channel. Then, the total XCI is the sum of the formally identical, albeit quantitatively different, contributions of each of the INT channels present in the WDM comb. In other words, the total PSD of XCI on the CUT is the sum of the PSDs generated due to each INT.

### A. The XPM approximation [24] to XCI

We started out from the formula provided in [24] in summation form, which the authors define as 'XPM'. We re-wrote it in integral dual-polarization form and in such a way as to make it represent the NLI power spectral density (PSD) emerging at a generic frequency $f$ within the CUT. It is:

$$G_{\text{XPM}}(f) = P_{\text{CUT}} P_{\text{INT}}^2 \left[ \kappa_{11}(f) + \Phi_b \, \kappa_{12}(f) \right] \tag{14}$$

where:

$$\Phi_b = \frac{\text{E}\left\{|b|^4\right\}}{\text{E}^2\left\{|b|^2\right\}} - 2 \tag{15}$$

$$\kappa_{11}(f) = \frac{32}{27} R_s^3 \int\limits_{-R_s/2}^{+R_s/2} df_1 \int\limits_{f_c-R_s/2}^{f_c+R_s/2} df_2$$

$$\left| s_{\text{CUT}}(f_1) \right|^2 \left| s_{\text{INT}}(f_2) \right|^2 \left| s_{\text{INT}}(f_1 + f_2 - f) \right|^2 \left| \mu(f_1, f_2, f) \right|^2 \tag{16}$$

$$\kappa_{12}(f) = \frac{80}{81} R_s^2 \int\limits_{-R_s/2}^{+R_s/2} df_1 \int\limits_{f_c-R_s/2}^{f_c+R_s/2} df_2 \int\limits_{f_c-R_s/2}^{f_c+R_s/2} df_2' \left| s_{\text{CUT}}(f_1) \right|^2$$

$$s_{\text{INT}}(f_2) s_{\text{INT}}^*(f_2') s_{\text{INT}}^*(f_1 + f_2 - f) s_{\text{INT}}(f_1 + f_2' - f) \, \mu(f_1, f_2, f) \mu^*(f_1, f_2', f) \tag{17}$$

As argued in [24], the $\kappa_{11}(f)$ term corresponds to a GN-model-like contribution, that is, it assumes signal Gaussianity. Instead, $\kappa_{12}(f)$ represents a correction that takes into account the non-Gaussianity of the transmitted signal. As said, these formulas account for a single INT channel. Considering a WDM system, the same calculations shown above must be repeated for each INT channel and the results summed together.

Note that in [24] XPM is not proposed as a partial contribution to NLI, but as an overall NLI estimator, accurate enough to represent the whole non-linearity affecting the CUT (excluding SCI). In the next subsection we will discuss this claim.

### B.  The overall XCI

Equation (14), derived from [24], neglects various XCI contributions arising when the INT channel is directly adjacent to the CUT. To provide a graphical intuitive description of what is left out, in Fig. 2 we show a plot of the domains in the $\left[ f_1, f_2 \right]$ plane where integration takes place for the $\kappa_{11}(f)$ and $\kappa_{12}(f)$ contributions. In the following paragraph we comment on why it is possible to discuss the integration domain of $\kappa_{12}(f)$ on the plane $\left[ f_1, f_2 \right]$, despite the fact that $\kappa_{12}(f)$ involves integration over three variables: $f_1, f_2, f_2'$.

As pointed out in [11], each point of the $\left[ f_1, f_2 \right]$ plane represents a triple of frequencies, namely $\left( f_1, f_2, f_3 \right)$, that produce a 'FWM' beat at frequency $f$. They obey the fixed relation $f_3 = f_1 + f_2 - f$. The 'elementary' NLI contributions, that are then integrated in the EGN formulas to provide the total NLI, arise each from *two* triples: $\left( f_1, f_2, f_3 \right)$ and $\left( f_1', f_2', f_3' \right)$, both producing a FWM contribution at the *same* frequency $f = \left( f_1 + f_2 - f_3 \right) = \left( f_1' + f_2' - f_3' \right)$. There are other constraints that relate the pairs of triples, which depend on the statistical features of the signal. For more details, see the appendices of this paper. It turns out that all different NLI contributions can be fully categorized just based on properly dividing the $\left[ f_1, f_2 \right]$ plane into integration regions where the $\left( f_1, f_2, f_3 \right)$ triples are located. This is because, if the subdivision is done correctly, the $\left( f_1', f_2', f_3' \right)$ triples that interact with each

$(f_1, f_2, f_3)$ triple, for a specific type of NLI, are bound to originate from the same region of the $[f_1', f_2']$ plane as that of the $[f_1, f_2]$ plane where $(f_1, f_2, f_3)$ originates. In other words, discussing the integration regions in $[f_1, f_2]$ is enough, because for each region in the $[f_1, f_2]$ plane the relevant region in the $[f_1', f_2']$ plane is the same, in a one-to-one correspondence.

The example of Fig. 2 considers XCI due to a single INT channel adjacent to the CUT, placed at higher frequency than the CUT, and assumes $f = 0$. The XPM formulas reported in [24], and hence Eq. (14), take into account the two X1 domains only. They neglect X2, X3 and X4. The complete XCI formulas that take all regions X1-X4 into account, are:

$$
\begin{aligned}
G_{\mathrm{XCI}}^{\mathrm{EGN}}(f) = & P_{\mathrm{CUT}} P_{\mathrm{INT}}^2 \left[ \kappa_{11}(f) + \Phi_b \ \kappa_{12}(f) \right] + \\
& P_{\mathrm{CUT}}^2 P_{\mathrm{INT}} \left[ \kappa_{21}(f) + \Phi_a \ \kappa_{22}(f) \right] + \\
& P_{\mathrm{CUT}}^2 P_{\mathrm{INT}} \left[ \kappa_{31}(f) + \Phi_a \ \kappa_{32}(f) \right] + \\
& P_{\mathrm{INT}}^3 \left[ \kappa_{41}(f) + \Phi_b \ \kappa_{42}(f) + \Psi_b \ \kappa_{43}(f) \right]
\end{aligned}
\tag{18}
$$

where:

$$
\Phi_a = \frac{\mathrm{E}\left\{|a|^4\right\}}{\mathrm{E}^2\left\{|a|^2\right\}} - 2 \quad , \quad \Phi_b = \frac{\mathrm{E}\left\{|b|^4\right\}}{\mathrm{E}^2\left\{|b|^2\right\}} - 2 \quad , \quad \Psi_b = \frac{\mathrm{E}\left\{|b|^6\right\}}{\mathrm{E}^3\left\{|b|^2\right\}} - 9 \frac{\mathrm{E}\left\{|b|^4\right\}}{\mathrm{E}^2\left\{|b|^2\right\}} + 12
\tag{19}
$$

The functions $\kappa_{mn}(f)$ are shown in Appendix B. Their derivation can be found in Appendix C. The index $m$ refers to the domain number according to Fig. 2. The GN-model part of XCI, $G_{\mathrm{XCI}}^{\mathrm{GN}}(f)$, stems from the $\kappa_{m1}(f)$ functions, with $m = 1 \ldots 4$. All other functions generate the correction part $G_{\mathrm{XCI}}^{\mathrm{corr}}(f)$. Similar to the SCI formula, when the correction contributions are addressed, both 4th order ($\Phi_a$ and $\Phi_b$) and 6th order ($\Psi_b$) moments of the transmitted symbol sequences must be considered, whereas in the XPM approximation only 4th order moments are involved.

Note the important circumstance that the XCI domains X2-X4 are non-empty as long as the INT channel adjacent to the CUT is not too far from the CUT, depending on the value of both $f$ and $|f_c|$. All three regions X2-X4 completely disappear when $|f_c| \geq 2R_s$, for any value of $f$ in the CUT band. This is automatically accounted for in Eq. (18), which can hence be considered a generalized complete formula for XCI, valid for channels adjacent to the CUT but also for non-adjacent channels, placed at any frequency distance from the CUT.

Even though the extra XCI X2-X4 regions appear only for the two channels adjacent to the CUT, they may contribute substantially to the overall NLI variance, depending on link and system parameters, so that disregarding them may lead to non-negligible error. This is due to the fact that these regions are relatively close to the origin of the $[f_1, f_2]$, where the $\mu$ integrand factors are maximum (see [11] for more details).

We investigated this matter by looking at the XCI normalized variance $\eta_{\mathrm{XCI}}$ defined as follows:

$$
\eta_{\mathrm{XCI}} = P_{\mathrm{ch}}^{-3} \int_{-R_s/2}^{R_s/2} G_{\mathrm{XCI}}^{\mathrm{EGN}}(f) df
\tag{20}
$$

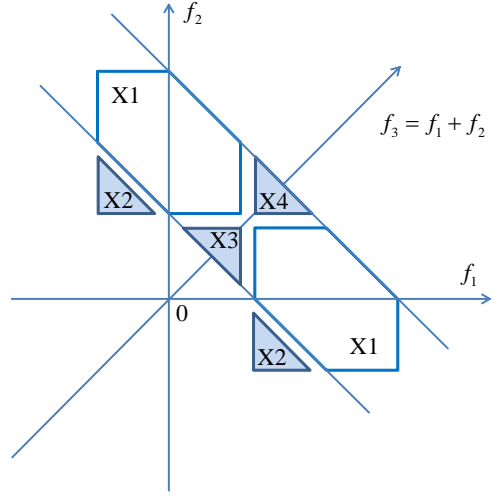

Fig. 2: Integration regions to obtain the power spectrum of XCI, $G_{\mathrm{XCI}}^{\mathrm{GN}}(f)$, at $f = 0$ (i.e., at the center of CUT), due to a single adjacent INT channel, assuming that its center frequency is slightly higher than the symbol rate. The XPM approximation [24] of Eq. (14) considers the X1 regions only. The full XCI formula of Eq. (18) accounts for all X1-X4 regions.

with $G_{\mathrm{XCI}}(f)$ given by Eq. (18). This parameter collects the total XCI noise spectrally located over the CUT, normalized so that $\eta_{\mathrm{XCI}}$ itself does not depend on launch power. Note that for simplicity we assume here:

$$P_{\mathrm{ch}} = P_{\mathrm{INT}} = P_{\mathrm{CUT}} \tag{21}$$

We calculated $\eta_{\mathrm{XCI}}$ for the same system addressed in Sect. 3 for SCI. The only difference is that now the system has 3 channels, with the CUT as the center channel. The channel spacing is $\Delta f = 33.6\,\mathrm{GHz}$. For the same system we also calculated $\eta_{\mathrm{XPM}}$, defined as:

$$\eta_{\mathrm{XPM}} = P_{\mathrm{ch}}^{-3} \int_{-R_s/2}^{R_s/2} G_{\mathrm{XPM}}(f)\,df \tag{22}$$

with $G_{\mathrm{XPM}}(f)$ given by Eq. (14).

Finally, still for the same system, we simultatively estimated the overall non-linearity, with single-channel effects removed. We did this because we wanted to see whether either XPM, or XCI, could be considered good approximations to the overall NLI produced in the link, once SCI is taken out. To remove SCI from the simulation results, we simulated both the CUT alone and the CUT with the two INT channels. Then we subtracted the former simulation result from the latter at the field level, thus ideally freeing the CUT completely from single-channel effects while leaving in all other non-linearity (XCI and MCI).

Figure 3(a) shows the XPM approximation $\eta_{\mathrm{XPM}}$ of [24] provided by Eq. (22) as a magenta solid line. The green solid line represents $\eta_{\mathrm{XCI}}$ given by the EGN-model Eq. (20). The red dashed curve represents the simulation result accounting for all NLI except SCI. All curves are represented as a function of the number of spans, up to 50.

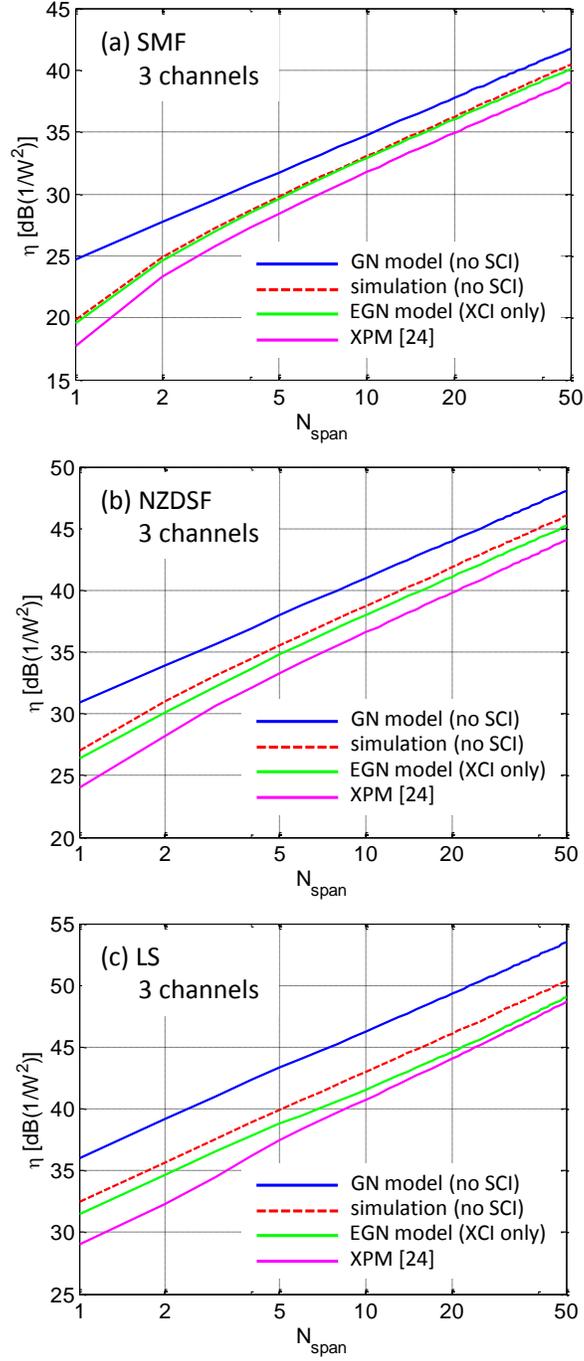

Fig. 3: Plot of normalized non-linearity coefficient $\eta$ vs. number of spans in the link, assuming *three* PM-QPSK channels over (from top to bottom) SMF, NZDSF and LS, with span length 100 km. The CUT is the center channel. The spacing is 1.05 times the symbol rate. Red dashed line: simulation, with single-channel non-linearity (SCI) removed. Blue solid line: GN-model without SCI. Magenta solid line: the XPM approximation $\eta_{XPM}$ of [24] (Eq. (22) of this paper). Green solid line: $\eta_{XCI}$ estimated through the EGN-model (Eq. (20)).

This may seem a large number of spans but the reach of the simulated system, assuming SMF, conventional EDFA amplification with realistic noise figure (5-6 dB) and a realistic FEC BER threshold of about $10^{-2}$, is indeed on the order of 50 spans. The figure shows that in this specific scenario the XPM approximation $\eta_{\mathrm{XPM}}$ of [24] underestimates the simulated NLI by about 1.4 dB. XCI $\eta_{\mathrm{XCI}}$ reduces such error to less than 0.4 dB throughout the plot. The GN-model starts out with a large 5-dB overestimation error, which gradually tapers down to about 1.3 dB at 50 spans.

In Fig. 3(b), we show a similar plot, this time for NZDSF. Above 5 spans, $\eta_{\mathrm{XPM}}$ of Eq. (22) underestimates NLI by about 2 dB whereas the GN-model overestimates it by about the same amount. These gaps are substantially wider than in the SMF case. Interestingly, a 0.8 dB gap is now also present between the simulation results and $\eta_{\mathrm{XCI}}$. This suggests that some NLI contributions are missing, i.e., the XCI component is not sufficiently representative of the overall NLI (excluding SCI).

A similar situation is also seen in Fig. 3(c), for the very low-dispersion scenario of LS fiber, with the interesting aspect that both XPM and XCI show a substantial underestimation error (1.7 and 1.3 dB, respectively) for a large number of spans. The GN-model clearly does not cope well with ultra-low dispersion fibers, showing a wide overestimation error of about 3.2 dB across all spans.

In conclusion, Figs. 3(a)-3(c) show that the XCI component of NLI may be sufficiently representative of all NLI (excluding SCI) only over high-dispersion fibers. On low-dispersion fibers part of NLI is clearly missing. In these specific examples, XPM is not representative of all of NLI and not even of XCI alone.

These results compellingly suggest that a complete model for NLI must include MCI as well. We introduce it in the next section. As a last remark, we point out that for larger values of the channel spacing $|f_c|$, a smaller gap can be expected between simulations and XPM, especially over SMF. Also, for $|f_c| \geq 2R_s$ XPM and XCI would coincide due to the vanishing of the X2-X4 regions.

## 5.  Multi-channel interference (MCI)

MCI can be thought of as typically being weaker than either SCI or XCI, because it arises on regions of the $[f_1, f_2]$ plane where the link function $\mu$ has a smaller magnitude than over the regions generating XCI and SCI. To provide an intuitive pictorial description of this circumstance, we show in Fig. 4 the integration regions arising in the plane $[f_1, f_2]$ when calculating the overall NLI PSD at the center of the CUT, i.e., $G_{\mathrm{NLI}}(0)$, for a three-channel example similar to the test PM-QPSK system of the previous section. The center region is SCI, the blue regions are XCI and the pink/red ones are MCI. Each point in these regions contributes to NLI, but it is weighed through the factors $\mu$ appearing in the integrals. These factors peak at the origin and along the $[f_1, f_2]$ plane axes. The larger the fiber dispersion is, the faster the decay of the $\mu$ factors away from such maxima. However, when dispersion is relatively low, such as with TrueWave RS or LS fibers, the decay of $\mu$ is much slower and MCI is not negligible, as the results of the previous section suggest.

Note also that when $G_{\mathrm{NLI}}(f)$ is evaluated at a frequency $f$ which is different than 0, the overall picture changes quite significantly. In particular, for $f \approx \pm R_s/2$ (values that correspond to the cut-off edges for a filter matched to a pulse $s_{\mathrm{CUT}}(f)$ with rectangular spectrum) some of the MCI integration regions come close to where the $\mu$'s are at their maxima. This case is exemplified in Fig. 5, which depicts the integration regions for $f = R_s/2$. The lower M0 and especially the lower M1 region are next to the $\mu$'s maxima, whose location has shifted away from the $[f_1, f_2]$ axes and now occurs at the red dashed axes. In this situation, MCI may therefore contribute substantially.

The MCI formulas for the red regions of Fig. 4 and Fig. 5 are:

$$G_{\text{MCI}}^{\text{EGN}}(f) = P_{\text{CUT}} P_{\text{INT,1}} P_{\text{INT,-1}} \kappa_{\text{M0}}(f) + P_{\text{INT,1}}^2 P_{\text{INT,-1}} \left[ \kappa_{\text{M1,1}}(f) + \Phi_b \, \kappa_{\text{M1,2}}(f) \right] \qquad (23)$$

where:

$$\kappa_{\text{M0}}(f) = 2 \cdot \frac{16}{27} R_s^3 \int_{f_c - R_s/2}^{f_c + R_s/2} df_1 \int_{-f_c - R_s/2}^{-f_c + R_s/2} df_2$$
$$\left| s_{\text{INT,1}}(f_1) \right|^2 \left| s_{\text{INT,-1}}(f_2) \right|^2 \left| s_{\text{CUT}}(f_1 + f_2 - f) \right|^2 \left| \mu(f_1, f_2, f) \right|^2 \qquad (24)$$

$$\kappa_{\text{M1,1}}(f) = 4 \cdot \frac{16}{27} R_s^3 \int_{-f_c - R_s/2}^{-f_c + R_s/2} df_1 \int_{f_c - R_s/2}^{f_c + R_s/2} df_2$$
$$\left| s_{\text{INT,1}}(f_1) \right|^2 \left| s_{\text{INT,1}}(f_2) \right|^2 \left| s_{\text{INT,1}}(f_1 + f_2 - f) \right|^2 \left| \mu(f_1, f_2, f) \right|^2 \qquad (25)$$

$$\kappa_{\text{M1,2}}(f) = 2 \cdot \frac{80}{81} R_s^2 \int_{-f_c - R_s/2}^{-f_c + R_s/2} df_1 \int_{f_c - R_s/2}^{f_c + R_s/2} df_2 \int_{f_c - R_s/2}^{f_c + R_s/2} df_2' \left| s_{\text{INT,1}}(f_1) \right|^2 s_{\text{INT,1}}(f_2) s_{\text{INT,1}}^*(f_2')$$
$$s_{\text{INT,1}}^*(f_1 + f_2 - f) s_{\text{INT,1}}(f_1 + f_2' - f) \mu(f_1, f_2, f) \mu^*(f_1, f_2', f) \qquad (26)$$

The subscripts 'INT,-1' and 'INT,1' refer to the INT channel spectrally located, respectively, to the left (lower frequency) and to the right (higher frequency) of the CUT.

Interestingly, in the pink region M0, NLI is produced entirely according to the GN-model, through $\kappa_{\text{M0}}$. No correction term for signal non-Gaussian distribution is present there. In the red region M1, the induced MCI has instead a similar structure as XCI in the blue region X1. In particular, both a GN-model-like term $\kappa_{\text{M1,1}}$ and a correction term $\kappa_{\text{M1,2}}$ are present.

For the same system set-ups addressed in Sect. 4B we calculated $\eta_{\text{MCI}}$, defined as:

$$\eta_{\text{MCI}} = P_{\text{ch}}^{-3} \int_{-R_s/2}^{R_s/2} G_{\text{MCI}}^{\text{EGN}}(f) df \qquad (27)$$

with $G_{\text{MCI}}^{\text{EGN}}(f)$ given by Eq. (23). We then summed together the XCI and MCI contributions. We call the result 'XMCI' for brevity:

$$\eta_{\text{XMCI}} = \eta_{\text{XCI}} + \eta_{\text{MCI}} \qquad (28)$$

where $\eta_{\text{XCI}}$ is given by Eq. (20). The quantity $\eta_{\text{XMCI}}$ is the green solid line in Figs. 6(a)-6(c). All curves except the green solid one are the same as in Figs. 3(a)-3(c). Comparing the two sets of figures, we see that the gap that existed between XCI and simulations has now completely disappeared. The gap was therefore due to the missing MCI contributions. The accuracy of the EGN-model in estimating $\eta_{\text{XMCI}}$ is remarkable, for both SMF and NZDSF. A small error shows up for LS in the first few spans, which completely disappears along the link.

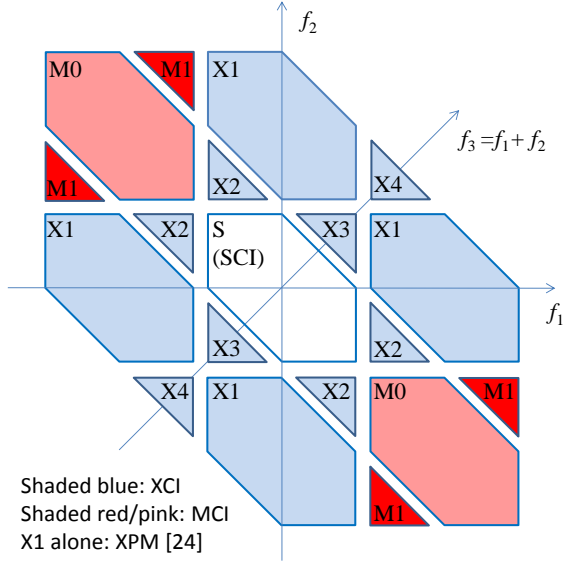

Fig. 4: Integration regions in the $\left[ f_1, f_2 \right]$ plane needed to obtain the power spectrum of NLI for $f = 0$, due to two adjacent INT channels with spacing slightly higher than the symbol rate. The full XCI formula of Eq. (20) accounts for all X1-X4 regions. The XPM approximation [24] (Eq. (22) here) considers the X1 regions only. SCI is the center region S. MCI is the red/pink regions. The M0 region has only the GN-model term, the red M1 ones have both the GN-model term and non-Gaussianity correction terms.

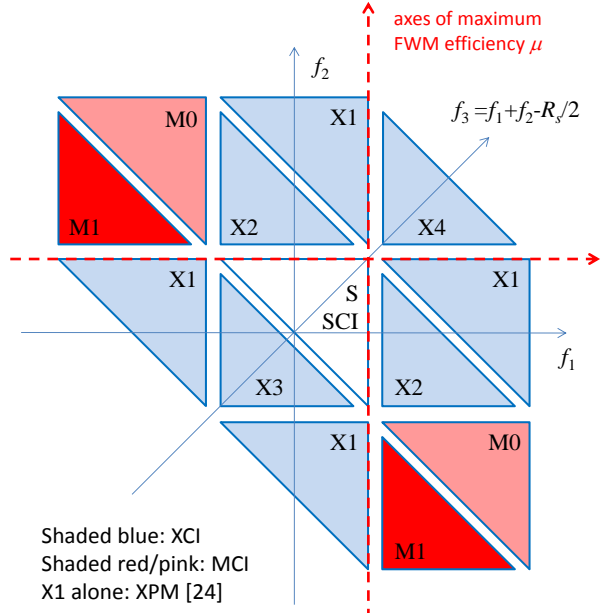

Fig. 5: Integration regions in the $\left[ f_1, f_2 \right]$ plane needed to obtain the power spectrum of NLI for $f = R_s / 2$, due to two adjacent INT channels with spacing slightly higher than the symbol rate. Notice that all regions change shape vs. Fig. 4. Also, the maximum FWM efficiency now falls on the translated red-dashed axes, which do not coincide with the $\left[ f_1, f_2 \right]$ axes. The lower M0 and M1 MCI regions are now close to such maxima.

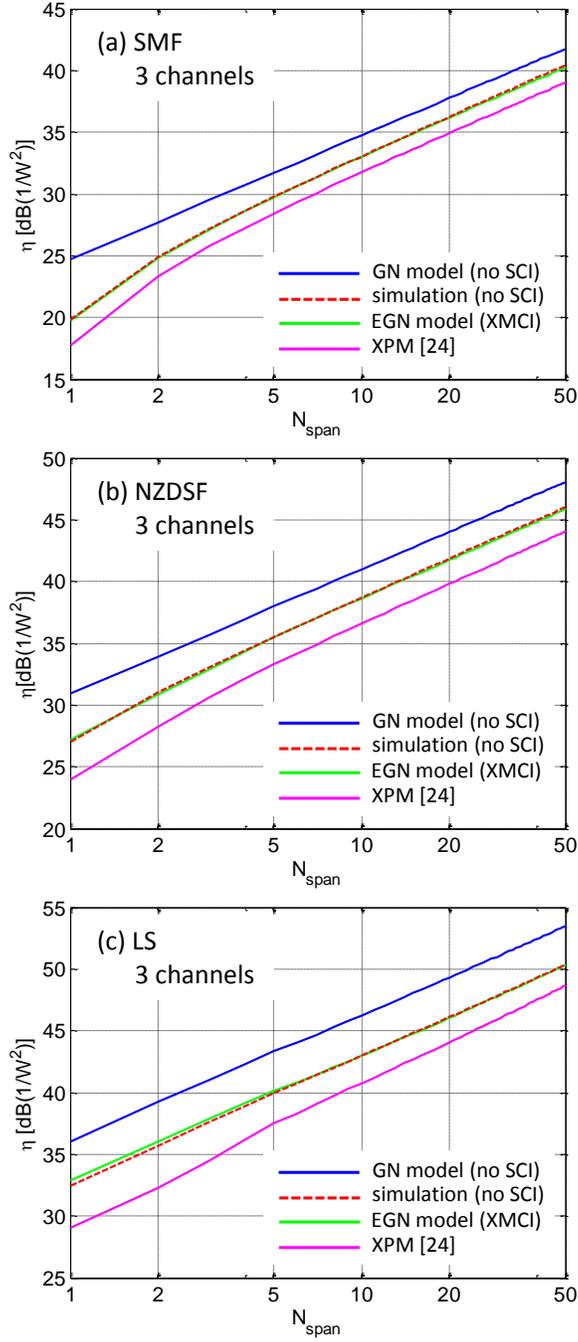

Fig. 6: Plot of normalized non-linearity coefficient $\eta$ vs. number of spans in the link, assuming *three* PM-QPSK channels over (from top to bottom) SMF, NZDSF and LS, with span length 100 km. The CUT is the center channel. The spacing is 1.05 times the symbol rate. Red dashed line: simulation, with single-channel non-linearity (SCI) removed. Blue solid line: GN-model without SCI. Magenta solid line: the XPM approximation $\eta_{XPM}$ of [24] (Eq. (22) of this paper). Green solid line: $\eta_{XMCI}$ (i.e., XCI+MCI) estimated through the EGN-model (Eq. (28)).

These results all assume just three channels. An interesting issue is whether the general picture shown in Figs. 6(a)-6(c) changes when going to a higher number of channels. One might wonder whether the extent and/or hierarchy of the gaps vs. simulation may change among curves, or whether the EGN-model might lose accuracy. This issue is dealt with in the next subsection, which also generalizes the MCI formulas to any number of WDM channels.

## A. MCI for any number of WDM channels

When more than three channels are present in the comb, the picture of the MCI integration regions becomes more complex. In Fig. 7 we show an example of a nine-channel quasi-Nyquist WDM system, assuming $f = 0$ for simplicity. The plot contains all possible types of MCI regions, together with those generated by SCI and XCI. Even going to a higher channel number than nine, no new region types are generated.

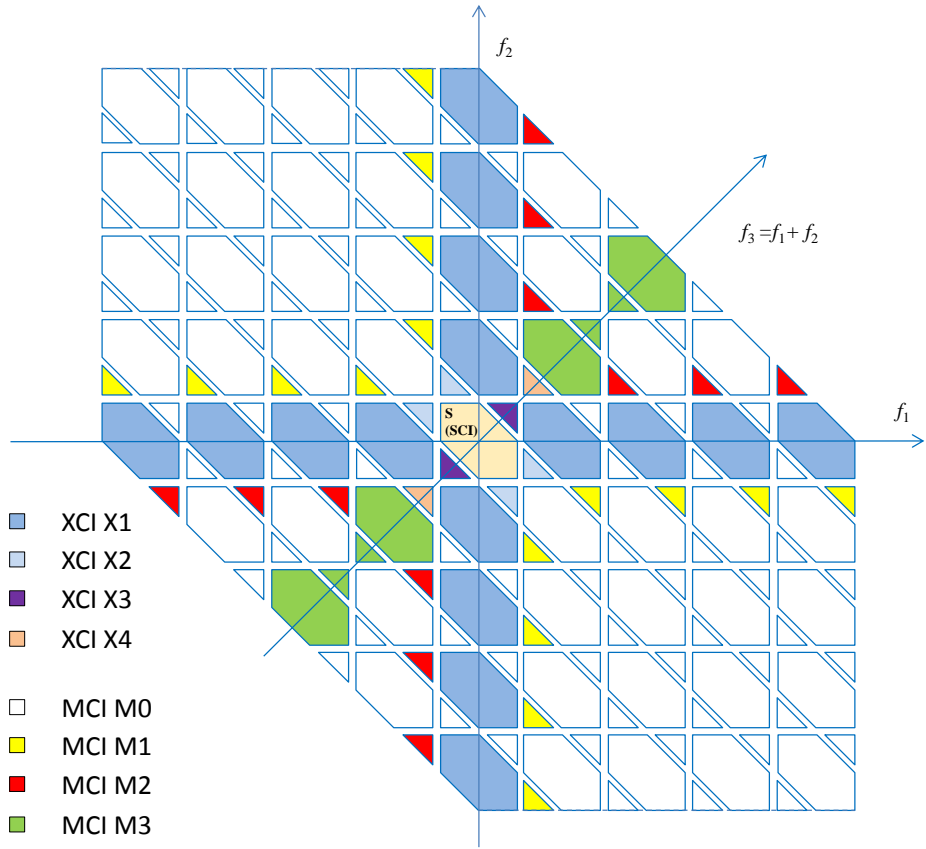

Fig. 7: Integration regions in the $\left[ f_1, f_2 \right]$ plane needed to obtain the power spectrum of NLI for $f = 0$, for a nine-channel WDM system with four left and four right INT channels adjacent to the CUT, with spacing slightly higher than the symbol rate. SCI is the center region. XCI and MCI regions are color-coded (see legend). The white-filled regions (all of type M0) have only the GN-model term, all others have both the GN-model term and one or more non-Gaussianity correction terms. Note that XPM amounts to the X1 regions only.

We generalized the MCI formulas to any number of channels, i.e., all four different MCI region types of Fig. 7 were addressed (see Appendix D). Such equations, together with the ones for SCI and XCI, make the overall EGN-model capable of dealing with any number of channels, for any type of NLI. Note that the MCI domains M1 and M2 are non-empty as long as the INT channel adjacent to the CUT is not too far from the CUT. Both regions M1 and M2 disappear when $|f_c| \geq 2R_s$, for any value of $f$ in the CUT band. This is automatically accounted for in the equations of Appendix D.

Using these general formulas, in Figs. 8(a)-8(c) we draw the same plot as Figs. 6(a)-6(c), except now *nine* WDM channels are present: the CUT and four adjacent INT channels on each side of the CUT. A comparison of the figures shows that, interestingly, the general picture is unchanged. The excellent accuracy of the EGN-model in estimating $\eta_{\text{XMCI}}$ is confirmed (green solid line) vs. simulations (red dashed), at this higher channel count too, for all fibers. The gap between simulations and the GN-model either slightly grows (for SMF) or is somewhat reduced. The gap between the XPM approximation and simulation decreases slightly for SMF but grows for NZDSF and quite substantially over LS (going from 1.3 to 3.1 dB).

## 6. Estimating System Performance

From the results of the previous sections, a rather compelling set of indications on the various addressed models emerges:

1. the EGN-model is very accurate in predicting XCI and MCI and quite accurate in predicting SCI too;
2. the XCI contributions of the X2-X4 regions and the MCI contributions to NLI may be important and cannot, in general, be neglected;
3. neither the GN-model, nor the XPM approximation to the EGN-model (SCI excluded) are accurate NLI estimators in any of the specific 3 or 9-channel examples addressed above.

In this section, we shift focus from the characterization of NLI accumulation along the link to system analysis. In fact, the main declared goal of the GN-model has always been that of providing a practical tool for realistic system performance prediction. Here, we present a comparison of the accuracy of the GN-model and of the EGN-model in predicting maximum system reach.

The systems that we tested are identical to those described in [14], Sect. V. Specifically, they are 15-channel WDM PM-QPSK, and PM-16QAM systems, running at 32 GBaud. The simulation technique is also similar to that of [14]. The simulated data length was 130,000 symbols. The target BERs were $1.7 \cdot 10^{-3}$ and $2 \cdot 10^{-3}$ respectively, found by assuming a $10^{-2}$ FEC threshold, decreased by 2 dB of realistic OSNR system margin. We considered the following channel spacings: 33.6, 35, 40, 45 and 50 GHz. The spectrum was root-raised-cosine with roll-off 0.05. EDFA amplification was assumed, with 5 dB noise figure. Single-channel non-linear effects were *not* removed from the simulation. The considered fibers were: SMF, NZDSF and LS, with same parameters as listed in Sect. 3, with the exception of the SMF loss that was $\alpha_{\text{dB}}$ =0.2 [dB/km] rather than 0.22. In addition, we also considered PSCF with the following parameters: $D$ =20.1 [ps/(nm km)], $\gamma$ =0.8 [1/(W km)], $\alpha_{\text{dB}}$ =0.17 [dB/km]. For more details on the simulation set-up and techniques, see [14], Sect. V.

Figure 9 shows a plot of maximum system reach vs. channel spacing. Squares are simulation results. The dashed line is the GN-model and the solid line is the EGN-model. Note that lines are just visual aids. The actually calculated data points are the filled circles. The GN-model underestimates the maximum reach by 0.3-0.6 dB over PSCF, SMF and NZDSF, in agreement with [14,21]. The error goes up to 0.8 dB in the case of the very low dispersion and high non-linearity LS fiber.

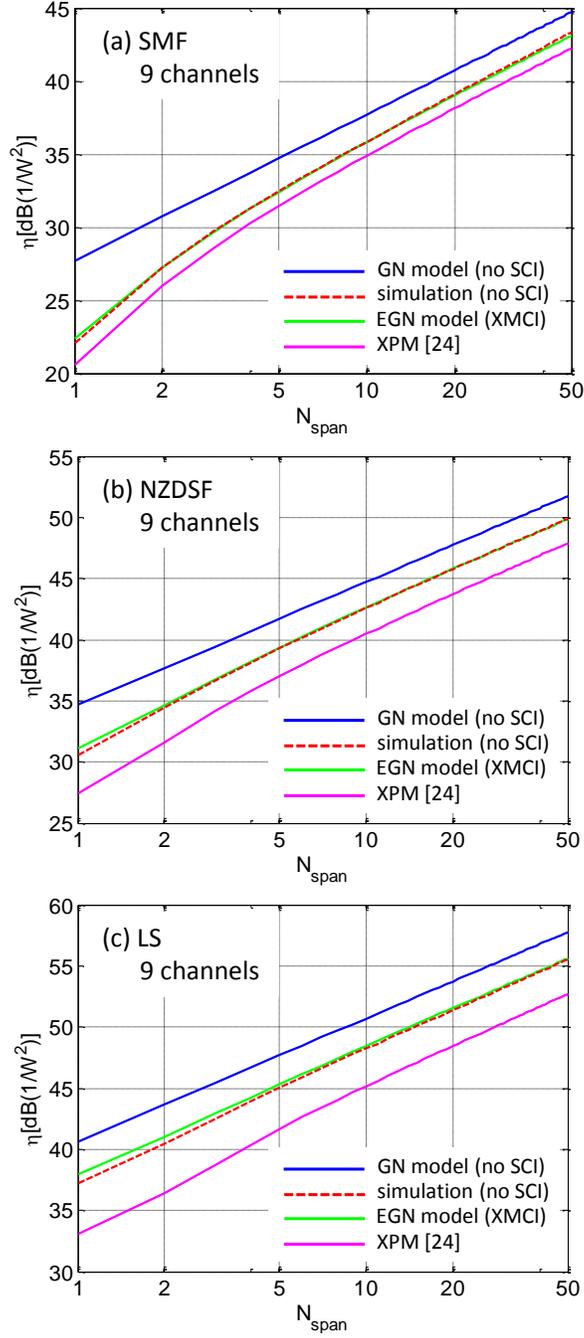

Fig. 8: Plot of normalized non-linearity coefficient $\eta$ vs. number of spans in the link, assuming *nine* PM-QPSK channels over (from top to bottom) SMF, NZDSF and LS, with span length 100 km. The CUT is the center channel. The spacing is 1.05 times the symbol rate. Red dashed line: simulation, with single-channel non-linearity (SCI) removed. Blue solid line: GN-model without SCI. Magenta solid line: the XPM approximation $\eta_{XPM}$ of [24] (Eq. (22)). Green solid line: $\eta_{XMCI}$ (i.e., XCI+MCI) estimated through the EGN-model (Eq. (28)).

These errors are in line with the typical amount of NLI overestimation by the GN-model that emerges from the previous sections, when taking into account that its impact on maximum reach error is downscaled by a factor 1/3, dB over dB [11,14].

With all fibers and spacings, the EGN-model provides very good accuracy, completely removing the underestimation error incurred by the GN-model. The error is less than 0.2 dB across all system configurations. At this error level, it is difficult to attribute it to either model inaccuracy or Monte-Carlo uncertainty.

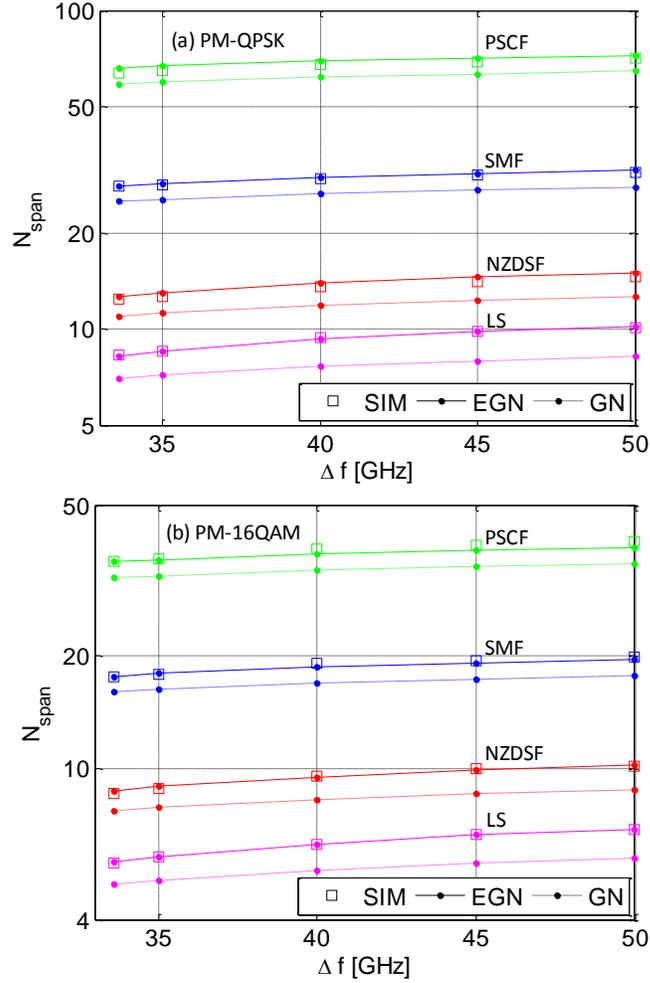

Fig. 9: Plot of maximum system reach for 15-channel PM-QPSK and PM-16QAM systems at 32 GBaud, vs. channel spacing, over four different fiber types: PSCF, SMF, NZDSF and LS. The span length is 120 km for PM-QPSK and 85 km for PM-16QAM. Small filled circles: analytical predictions. Square hollow markers: simulations. Lines were added to connect analytical points as a visual aid. Dashed line: GN-model. Solid line: EGN-model.

We would like to point out that a slight difference, on the order of small fractions of a dB, is visible between some of the system results shown in [14], and the ones reported here in Fig. 9. They are due to two circumstances. First, in [14] the local-white-noise approximation was used in the calculation of NLI using the GN-model. Such approximation consists of assuming that the NLI spectrum is essentially flat over the bandwidth of the channel under test. Here, the non-flatness of the NLI spectrum was fully taken into account when plotting all the figures in this paper. Specifically regarding Fig. 9, the difference between taking and not taking the non-flat NLI spectrum into account causes an upshift of the analytical curves ranging between 0.05 dB for $\Delta f$ =33.6 GHz and 0.15 dB for $\Delta f$ = 50 GHz. As a result, the GN-model prediction here is different from [14] by this much.

A second difference with [14] is that the simulations there, for the sake of full realism, were run with ASE noise added in-line along the link. Here, we want to carefully validate a model that neglects the interaction of in-line ASE noise with non-linearity, so we added all ASE noise at the end of the link. The effect is that all simulative PM-QPSK results are pulled up here by about 0.15 dB on average. The effect on PM-16QAM is almost negligible (less than 0.05 dB), because PM-16QAM requires a much higher OSNR at the receiver and hence much less ASE noise is present along the link than for PM-QPSK.

We feel that neither of these small differences with respect to [14] changes the essence of the results shown either here or in [14].

## 7. Discussion

In this section we discuss various issues related to the GN and EGN-models: accuracy vs. computational effort, non-linear phase noise and the comparison of model results with experiments.

### A. Accuracy vs. computational effort

The GN-model overestimates NLI. The amount of overestimation is large in the first spans (several dB's) but it abates along the link. When looked at for a number of spans that is close to the maximum reach, the error on NLI power estimation is typically 1 to 2 dB, depending on fiber type, modulation format and span length, for realistic systems. Larger errors can be found by pushing the system parameters to extremes, such as single-polarization, lossless fiber (or ideal distributed amplification) or very short spans.

The GN-model errors in NLI power estimation in turn lead to about 0.3-0.6 dB of error on the prediction of the maximum reach or of the optimum launch power, for typical realistic systems. This error may or may not be acceptable, depending on applications, but is guaranteed to be conservative (i.e., reach is underestimated) for PM-QAM systems. When such error is not acceptable, the EGN-model can be used, which is capable of providing very accurate estimates of NLI variance at any number of spans along the link, potentially for any format and system set of parameters.

The results of Fig. 9 contain both simulations and analytical calculations. The simulations required a large CPU effort, due to our will to impose very strict accuracy constraints. Please see [14] for a description of simulation accuracy settings. As a whole, the simulated points populating Fig. 9 required several months of equivalent single-core CPU (PC-type) time. This should not surprise, since accurately finding maximum reach by simulation requires demodulating the signal at multiple spans and also scanning numerous launch powers at small steps. For each launch power, an entirely new simulation must be run. Some optimizations are possible but the overall burden is massive.

The EGN-model calculations needed to generate the corresponding data points, thanks to various optimizations, were trimmed down to about 15 days of total single-core CPU time. One key factor contributing to reducing the computational effort of the EGN-model is the fact that, even when three or more nested integrals are present in any of the NLI contributions, the actual complexity is always equivalent to a double integral. This aspect is explained in Appendix F. Another important speed-up circumstance vs. simulations is that the model calculations do not need to be run at different launch powers. Once the normalized coefficient $\eta_{NLI}$ has been estimated, NLI can be extrapolated to any power by simply scaling it analytically by $P_{ch}^3$. One circumstance acting against model calculations efficiency is however that if very high accuracy is needed, the NLI white-noise approximation used for instance in [14] must be avoided. This entails evaluating the NLI PSD (essentially $\eta_{\mathrm{NLI}}$) at many frequencies inside the CUT bandwidth (32 GHz in our case) and then averaging them. We used a step of 1 GHz, which we found sufficient.

Despite forgoing the white-noise approximation, the EGN-model CPU gain vs. Monte-Carlo simulations was still quantifiable as a factor of 10-20. It should however be mentioned that we were conservative as to setting the integration parameters for accuracy. Also, we think the efficiency of our code could be improved upon. As a result, we feel that it should be possible to push the previously mentioned speed-up factor to at least 20-40. This factor is significant. It is however not significant enough to make the EGN-model a real-time tool for quick system optimization. We should also point out that not even the GN-model can be considered a real-time tool, as the speed-up of the GN-model vs. the EGN-model is only about another factor of 5-10, insufficient for real-time use.

The fastest GN-related model available is the *incoherent* GN-model, whose accuracy was shown to typically appear to be even better than the GN-model [14]. This is somewhat surprising, since the incoherent GN-model is derived from the GN-model by making one further approximation, namely that the NLI produced in each span sums up incoherently (that is, in power) at the receiver [6,7,11,14]. However, as explained in [14], it benefits from an error cancellation circumstance. This means that, while the GN-model produces a guaranteed lower bound to the maximum reach, the incoherent GN-model can be either pessimistic or optimistic. On the other hand, its speed of computation is 10-20 times faster than the GN-model, and another order of magnitude can be gained if the white noise assumption is used. In essence, the incoherent GN-model is so far the fastest tool, and essentially a real-time tool, for system performance assessment. On the other hand, caution must be used and its limitations must be fully understood to use it properly. Its margin of error can potentially be substantial, although so far, in the context of many validation campaigns using realistic system parameters [6,7,14], it has been consistently found to be rather accurate. It should also be mentioned that a number of closed-form or quasi-closed form analytical solutions have also been worked out both for the GN-model and the incoherent GN-model [11,28,30], which clearly reduce complexity to almost negligible levels, at the cost of some potential loss of accuracy.

The best of all options would arguably be that of finding a tool with a similar complexity as the incoherent GN-model, whose accuracy would however not rest on an error cancellation, but on firm theoretical ground. A first promising attempt towards this direction, based on an analytical closed-form approximation to the EGN-model, is reported in [27].

Overall, an array of analytical tools are already available for the system designer, with different degrees of complexity and accuracy that can be tailored to specific needs. Trade-offs between accuracy and complexity can already be addressed with numerous options at hand.

*B. Non-linear phase noise*

As mentioned in the introduction, one of the assumptions used by the GN-model, as well as by most prior non-linearity models, is that of NLI being approximately Gaussian and additive, so that its system impact can be assessed simply by summing its variance to that of ASE noise. This assumption was challenged in [24] and [25]. The claim of [24] is that a very substantial part of the XCI contribution to NLI is in fact *phase noise* and hence non-additive. In addition, such phase noise appears to have a very long correlation time, on the order of tens or even hundreds of symbols.

The presence of a non-linear noise component with very long correlation time had first been pointed out in [26], there too attributed to 'cross-phase modulation'. The correlation results in [24] actually agree well with those found earlier in [26]. Both papers, however, concentrate on a single-polarization, lossless fiber scenario to assess the strength of the long-correlated phase-noise component of NLI. In that idealized context, the phase noise component may indeed turn out to be very large.

Clearly, the assumption of NLI being Gaussian and additive is an approximation. The key problem is whether such approximation is good enough for the purpose it was made. Such purpose was to make it possible to assess system performance by simply adding the ASE noise and the NLI variance at the denominator of a 'non-linear' effective optical signal-to-noise-ratio. In our opinion, the results presented in Fig. 9, where the maximum system reach was predicted with a very high level of accuracy using the EGN-model together with the additive-Gaussian NLI approximation, represent strong evidence that such approximation is quite adequate for dealing with practical system scenarios, when maximum system performance is investigated. In a separate forthcoming paper, we report on a specific in-depth investigation of the extent and features of non-linear phase-noise in practical links and on its impact on performance prediction.

*C. Modeling vs. actual systems and networks*

As shown, the EGN-model provides much better accuracy than the GN-model in predicting the span-by-span accumulation of NLI noise. The differences between the GN and EGN-model are due to the removal of the signal Gaussian distribution assumption. However, the reason why this removal impacts significantly the accumulation of NLI is not straightforward.

In uncompensated systems the signal does get substantially spread out due to dispersion and rather quickly takes on an approximately Gaussian distribution. Nonetheless, a residual dependence among the random variables appearing in the Fourier transform of each single channel (the $\nu_n$'s in Appendix A) survives the dispersive effect, eventually causing a reduction in the amount of NLI produced in the link even at large span count, as shown for instance in the NLI accumulation plots in this paper.

This effect shows up mathematically under various implied assumptions. One foundational assumption is that propagation is modeled through the Manakov equation. Another key assumption is that the channels travel together from input to output. A third one, is that ASE noise does not significantly impact non-linearity generation.

All three of these assumptions can be challenged, to various extents, depending on system environment and link parameters. For instance, already in current networks, and increasingly so in future ones, the WDM channels are routed in arbitrary ways along the links so that a given channel may change its neighbor interfering channels more than once along its path. This may weaken the high-coherence picture that is essential in producing the significant deviation of the EGN-model vs. the GN-model. Further research should be devoted to investigating the quantitative impact of this and other similar circumstances. In certain cases, the built-in conservative nature of the GN-model might turn out to constitute a safe margin towards possible random performance variability.

Regarding ASE noise, the effect of NLI produced by co-propagating ASE noise on system performance is small as long as the required OSNR at the receiver is relatively large. The trend towards using ever more complex FECs allowing operation at very low OSNRs suggests that this effect may become substantially more significant than the 0.15 dB assessed here for PM-QPSK (see Sect. 6). Both the GN and the EGN-model can be extended to analytically take it into account. This topic is however considered outside of the scope of this paper and left for future investigation.

These remarks recommend caution in the use of models when relating such models to the physical world. More in-depth comparison of model predictions and actual experimental results would in fact be desirable, to make sure that the many assumptions of all models pan out positively in the physical world.

## 8. Comments and Conclusion

In this paper we have provided the full set of formulas needed for a self-consistent complete EGN-model, derived using an extension of the procedure proposed in [24] to remove the signal Gaussianity assumption.

In detail, we have derived for the first time single-channel non-linearity formulas, which had not been addressed in [24]. We have also shown that the 'XPM' formulas proposed in [24] as an estimator for cross and multi-channel NLI (that is, of all NLI except single-channel) can substantially underestimate non-linear noise in certain scenarios, especially in systems with low-dispersion fibers, where additional NLI terms may be significant. We have provided the complete set of formulas describing all possible cross- and multi-channel interactions, and carefully validated them vs. simulations.

The EGN-model presented here exhibits the best predictive power so far, among the various approximate GN-related models available. This is not only true at a span count nearing maximum reach, but throughout the link. It shows no evident bias versus non-linearity over or underestimation. It can be used reliably at even ultra-low dispersion, such as over LS fibers. It can also potentially be used to study pre-compensation techniques and mixed fiber environments. Its effectiveness in these latter contexts will be investigated in a specifically devoted forthcoming paper.

Looking at the final EGN-model formulas, it is evident that the price to pay for its increased accuracy is increased complexity. In certain cases, the potential speed up vs. standard split-step simulations can be as low as just a factor of 20. A key objective for research in the near future is therefore that of trying to drastically reduce such complexity, perhaps by deriving from the EGN model suitable GN-model correction terms which permit to combine improved accuracy with reasonable complexity. A first result towards this goal is reported in [27].

**Appendix A: Derivation of the SCI formulas**

In frequency domain, the signal model for a single channel (dual polarization), can be written as [7]:

$$\vec{E}(f) = E_x(f)\hat{x} + E_y(f)\hat{y} \tag{29}$$

where:

$$E_x(f) = \sqrt{f_0} \sum_{n=-\infty}^{+\infty} v_{x,n} \delta(f - nf_0) \quad , \quad E_y(f) = \sqrt{f_0} \sum_{n=-\infty}^{+\infty} v_{y,n} \delta(f - nf_0) \tag{30}$$

The random variables $v_{x,n}$ and $v_{y,n}$ are defined similar to [7], Appendix B, Eq. (36):

$$v_{x,n} = \sqrt{f_0} s_{\text{CUT},x}(nf_0) \sum_{\omega=0}^{W-1} a_{x,\omega} e^{-j\frac{2\pi}{W}\omega n} \quad , \quad v_{y,n} = \sqrt{f_0} s_{\text{CUT},y}(nf_0) \sum_{\omega=0}^{W-1} a_{y,\omega} e^{-j\frac{2\pi}{W}\omega n} \tag{31}$$

where $W$ is an integer number that can be chosen to be arbitrarily large.

Using the Manakov equation, the Kerr term at the fiber input on the $x$ polarization can be written as (see [10], Eqs. (28)-(29), (33)-(34) and (75)):

$$Q_{\text{NLI},x}(0,f) = -j\gamma \frac{8}{9} f_0^{3/2} \sum_{i=-\infty}^{+\infty} \delta(f - if_0) \sum_{S_i} \left( v_{x,m} v_{x,n}^* v_{x,k} + v_{x,m} v_{y,n}^* v_{y,k} \right) \tag{32}$$

where:

$$S_i \equiv \{ (m,n,k) : (m-n+k)f_0 = if_0 \} \tag{33}$$

The SCI field on the $x$ polarization after a homogenous span of length $L_s$, can be written as:

$$E_{\text{SCI},x}(L_s,f) = f_0^{3/2} e^{-\alpha L_s} \sum_{i=-\infty}^{+\infty} \delta(f-if_0) \left[ -j \frac{8}{9} e^{-j2\beta_2\pi^2 i^2 f_0^2 L_s} \right.$$

$$\left. \sum_{m,n,k \in \mathbb{S}_i} \zeta(k,m,n) \cdot \left( v_{x,m} v_{x,n}^* v_{x,k} + v_{x,m} v_{y,n}^* v_{y,k} \right) \right] \tag{34}$$

where:

$$\zeta(k,m,n) = \gamma \frac{1 - e^{-2\alpha L_s} e^{j4\pi^2 \beta_2 f_0^2 (k-n)(m-n) L_s}}{2\alpha - j4\pi^2 \beta_2 f_0^2 (k-n)(m-n)} \tag{35}$$

By 'homogeneous span' we mean that the fiber parameters are constant over the length $L_s$.

The SCI PSD on the $x$ polarization at the output of the fiber span of length $L_s$ is then:

$$G_{\text{SCI},x}(f) = \frac{64}{81} f_0^3 e^{-2\alpha L_s} \sum_{i=-\infty}^{+\infty} \delta(f-if_0) \sum_{m,n,k \in \mathbb{S}_i} \sum_{m',n',k' \in \mathbb{S}_i} \zeta(k,m,n) \zeta^*(k',m',n')$$

$$\left[ \mathrm{E}\left\{ v_{x,m} v_{x,n}^* v_{x,k} v_{x,m'}^* v_{x,n'} v_{x,k'}^* \right\} + \mathrm{E}\left\{ v_{x,m} v_{x,n}^* v_{x,k} v_{x,m'}^* \right\} \mathrm{E}\left\{ v_{y,n'} v_{y,k'}^* \right\} \right.$$

$$\left. + \mathrm{E}\left\{ v_{x,m} v_{x,m'}^* v_{x,n'} v_{x,k'}^* \right\} \mathrm{E}\left\{ v_{y,n}^* v_{y,k} \right\} + \mathrm{E}\left\{ v_{x,m} v_{x,m'}^* \right\} \mathrm{E}\left\{ v_{y,n}^* v_{y,k} v_{y,n'} v_{y,k'}^* \right\} \right] \tag{36}$$

We calculate these expectations according to the formulas given in Appendix E, and rewrite the SCI PSD as:

$$G_{\text{SCI},x}(f) = \frac{64}{81} f_0^3 e^{-2\alpha L_s} \sum_{i=-\infty}^{+\infty} \delta(f-if_0) \sum_{m,n,k \in \mathbb{S}_i} \sum_{m',n',k' \in \mathbb{S}_i} \zeta(k,m,n) \zeta^*(k',m',n')$$

$$\left\{ 2R_s^3 \mathrm{E}^3\left\{|a_x|^2\right\} \mathcal{P}_1 \delta_{m-m'} \delta_{n'-n} \delta_{k-k'} + R_s^3 \mathrm{E}\left\{|a_x|^3\right\} \mathrm{E}^2\left\{|a_x|^2\right\} \mathcal{P}_2 \delta_{m-m'} \delta_{k-k'} \delta_{n'-n} \right.$$

$$+ R_s^2 f_0 \mathrm{E}^3\left\{|a_x|^3\right\} \left[ \mathrm{E}\left\{|a_x|^4\right\} / \mathrm{E}^2\left\{|a_x|^2\right\} - 2 \right] \mathcal{P}_1 \left( 4\delta_{m-m'} \delta_{k-n+n'-k'} + \delta_{n'-n} \delta_{m+k-m'-k'} \right)$$

$$+ R_s f_0^2 \mathrm{E}^3\left\{|a_x|^2\right\} \left[ \mathrm{E}\left\{|a_x|^6\right\} / \mathrm{E}^3\left\{|a_x|^2\right\} - 9 \mathrm{E}\left\{|a_x|^4\right\} / \mathrm{E}^2\left\{|a_x|^2\right\} + 12 \right] \mathcal{P}_1 \delta_{m-n+k-m'+n'-k'}$$

$$+ R_s^2 f_0 \mathrm{E}\left\{|a_x|^2\right\} \mathrm{E}^2\left\{|a_y|^2\right\} \cdot \left[ \mathrm{E}\left\{|a_y|^4\right\} / \mathrm{E}^2\left\{|a_y|^2\right\} - 2 \right] \mathcal{P}_2 \delta_{m-m'} \delta_{k-n+n'-k'} \right\} \tag{37}$$

where:

$$\mathcal{P}_1 = s_{\text{CUT},x}(mf_0) s_{\text{CUT},x}^*(nf_0) s_{\text{CUT},x}(kf_0) s_{\text{CUT},x}^*(m'f_0) s_{\text{CUT},x}(n'f_0) s_{\text{CUT},x}^*(k'f_0) \tag{38}$$

$$\mathcal{P}_2 = s_{\text{CUT},x}(mf_0) s_{\text{CUT},y}^*(nf_0) s_{\text{CUT},x}(kf_0) s_{\text{CUT},x}^*(m'f_0) s_{\text{CUT},y}(n'f_0) s_{\text{CUT},x}^*(k'f_0) \tag{39}$$

In addition, we remove the terms with $\{m=n$ or $k=n\}$ or $\{m'=n'$ or $k'=n'\}$ because they can be shown to contribute a frequency-flat, constant phase shift which has no detrimental effect on transmission [10] (Eqs. (33)-(45)), [12], [13], [24].

If we also assume that,

$$\mathrm{E}\left\{|a_x|^2\right\} = \mathrm{E}\left\{|a_y|^2\right\} = \frac{1}{2} \mathrm{E}\left\{|a|^2\right\} \quad , \quad s_{\text{CUT},x}(f) = s_{\text{CUT},y}(f) = s_{\text{CUT}}(f) \tag{40}$$

then we get the simplified expression:

$$G_{\text{SCI},x}(f) = \frac{8}{81} f_0^3 e^{-2\alpha L_s} \, \mathrm{E}^3\left\{|a|^2\right\} \sum_{i=-\infty}^{+\infty} \delta(f - i f_0)$$

$$\sum_{m,n,k \in \mathbb{S}_i} \sum_{m',n',k' \in \mathbb{S}_i} \zeta(k,m,n) \zeta^*(k',m',n') \left\{ 3R_s^3 \mathcal{P}_{\text{CUT}} \delta_{m-m'} \delta_{n'-n} \delta_{k-k'} \right.$$

$$+ R_s^2 f_0 \left[ \mathrm{E}\left\{|a|^4\right\} / \mathrm{E}^2\left\{|a|^2\right\} - 2 \right] \mathcal{P}_{\text{CUT}} \left( 5\delta_{m-m'} \delta_{k-n+n'-k'} + \delta_{n'-n} \delta_{m+k-m'-k'} \right)$$

$$\left. + R_s f_0^2 \left[ \mathrm{E}\left\{|a|^6\right\} / \mathrm{E}^3\left\{|a|^2\right\} - 9\mathrm{E}\left\{|a|^4\right\} / \mathrm{E}^2\left\{|a|^2\right\} + 12 \right] \mathcal{P}_{\text{CUT}} \delta_{m-n+k-m'+n'-k'} \right\}$$

$$\tag{41}$$

where:

$$\mathcal{P}_{\text{CUT}} = s_{\text{CUT}}(mf_0) s_{\text{CUT}}^*(nf_0) s_{\text{CUT}}(kf_0) s_{\text{CUT}}^*(m'f_0) s_{\text{CUT}}(n'f_0) s_{\text{CUT}}^*(k'f_0) \tag{42}$$

The contribution on the $y$ polarization is identical. Therefore, the total EGN-model SCI PSD is:

$$G_{\text{SCI}}(f) = \mathrm{E}^3\left\{|a|^2\right\} \left[ \chi_1(f) + \Phi_a \, \chi_2(f) + \Psi_a \, \chi_3(f) \right] \tag{43}$$

where:

$$\chi_1(f) = \frac{16}{81} f_0^3 e^{-2\alpha L_s} \sum_{i=-\infty}^{+\infty} \delta(f - i f_0)$$

$$\sum_{m,n,k \in \mathbb{S}_i} \sum_{m',n',k' \in \mathbb{S}_i} \zeta(k,m,n) \zeta^*(k',m',n') \cdot 3R_s^3 \mathcal{P}_{\text{CUT}} \delta_{m-m'} \delta_{n'-n} \delta_{k-k'}$$

$$= \frac{16}{27} f_0^3 e^{-2\alpha L_s} R_s^3 \sum_{i=-\infty}^{+\infty} \delta(f - i f_0) \sum_{m} \sum_{k} \left| s_{\text{CUT}}(mf_0) \right|^2$$

$$\left| s_{\text{CUT}}(kf_0) \right|^2 \left| s_{\text{CUT}}([m+k-i]f_0) \right|^2 \left| \zeta(m,k,i) \right|^2$$

$$\tag{44}$$

$$\chi_2(f) = \frac{16}{81} f_0^3 e^{-2\alpha L_s} \sum_{i=-\infty}^{+\infty} \delta(f - i f_0) \sum_{m,n,k \in \mathbb{S}_i} \sum_{m',n',k' \in \mathbb{S}_i}$$

$$\zeta(k,m,n) \zeta^*(k',m',n') \cdot R_s^2 f_0 \mathcal{P}_{\text{CUT}} \left( 5\delta_{m-m'} \delta_{k-n+n'-k'} + \delta_{n'-n} \delta_{m+k-m'-k'} \right)$$

$$= \frac{80}{81} f_0^4 e^{-2\alpha L_s} R_s^2 \sum_{i=-\infty}^{+\infty} \delta(f - i f_0) \sum_{m} \sum_{k} \sum_{k'} \left| s_{\text{CUT}}(mf_0) \right|^2 s_{\text{CUT}}(kf_0)$$

$$s_{\text{CUT}}^*(k'f_0) s_{\text{CUT}}^*([m+k-i]f_0) s_{\text{CUT}}([m+k'-i]f_0) \zeta(m,k,i) \zeta^*(m,k',i)$$

$$+ \frac{16}{81} f_0^4 e^{-2\alpha L_s} R_s^2 \sum_{i=-\infty}^{+\infty} \delta(f - i f_0) \sum_{m} \sum_{k} \sum_{k'} \left| s_{\text{CUT}}([m+k-i]f_0) \right|^2 s_{\text{CUT}}(mf_0)$$

$$s_{\text{CUT}}(kf_0) s_{\text{CUT}}^*(k'f_0) s_{\text{CUT}}^*([m+k-k']f_0) \zeta(m,k,i) \zeta^*(m+k-k',k',i)$$

$$\tag{45}$$

$$\chi_3(f) = \frac{16}{81} f_0^3 e^{-2\alpha L_s} \sum_{i=-\infty}^{+\infty} \delta(f - i f_0) \sum_{m,n,k \in \mathbb{S}_i} \sum_{m',n',k' \in \mathbb{S}_i}$$

$$\zeta(k,m,n) \zeta^*(k',m',n') \cdot R_s f_0^2 \mathcal{P}_{\text{CUT}} \delta_{m-n+k-m'+n'-k'}$$

$$= \frac{16}{81} f_0^5 e^{-2\alpha L_s} R_s \sum_{i=-\infty}^{+\infty} \delta(f - i f_0) \sum_{m} \sum_{k} \sum_{m'} \sum_{k'} s_{\text{CUT}}(mf_0) s_{\text{CUT}}^*([m+k-i]f_0)$$

$$s_{\text{CUT}}(kf_0) s_{\text{CUT}}^*(m'f_0) s_{\text{CUT}}([m'+k'-i]f_0) s_{\text{CUT}}^*(k'f_0) \zeta(m,k,i) \zeta^*(m',k',i)$$

$$\tag{46}$$

If identical spans of same fiber type are assumed, with lumped amplifiers exactly compensating for the loss of each span, the SCI PSD is:

$$G_{\text{SCI}}^{\text{EGN}}(f) = \text{E}^3\left\{|a|^2\right\}\left[\kappa_1(f) + \Phi_a\,\kappa_2(f) + \Psi_a\,\kappa_3(f)\right] \tag{47}$$

where:

$$\kappa_1(f) = \frac{16}{27} f_0^3 R_s^3 \sum_{i=-\infty}^{+\infty} \delta(f-if_0)\sum_m\sum_k$$
$$\left|s_{\text{CUT}}(mf_0)\right|^2\left|s_{\text{CUT}}(kf_0)\right|^2\left|s_{\text{CUT}}([m+k-i]f_0)\right|^2\left|\mu(m,k,i)\right|^2 \tag{48}$$

$$\kappa_2(f) = \frac{80}{81} f_0^4 R_s^2 \sum_{i=-\infty}^{+\infty} \delta(f-if_0)\sum_m\sum_k\sum_{k'}\left|s_{\text{CUT}}(mf_0)\right|^2 s_{\text{CUT}}(kf_0)$$
$$s_{\text{CUT}}^*([m+k-i]f_0)s_{\text{CUT}}([m+k'-i]f_0)s_{\text{CUT}}^*(k'f_0)\,\mu(m,k,i)\mu^*(m,k',i)$$
$$+\frac{16}{81} f_0^4 R_s^2 \sum_{i=-\infty}^{+\infty} \delta(f-if_0)\sum_m\sum_k\sum_{k'}\left|s_{\text{CUT}}([m+k-i]f_0)\right|^2$$
$$s_{\text{CUT}}(mf_0)s_{\text{CUT}}(kf_0)s_{\text{CUT}}^*([m+k-k']f_0)s_{\text{CUT}}^*(k'f_0)\,\mu(m,k,i)\mu^*(m+k-k',k',i) \tag{49}$$

$$\kappa_3(f) = \frac{16}{81} f_0^5 R_s \sum_{i=-\infty}^{+\infty} \delta(f-if_0)\sum_m\sum_k\sum_{m'}\sum_{k'}$$
$$s_{\text{CUT}}(mf_0)s_{\text{CUT}}^*([m+k-i]f_0)s_{\text{CUT}}(kf_0)s_{\text{CUT}}^*(m'f_0)$$
$$s_{\text{CUT}}([m'+k'-i]f_0)s_{\text{CUT}}^*(k'f_0)\,\mu(m,k,i)\mu^*(m',k',i) \tag{50}$$

The symbol $\mu$ is the 'link function', defined as:

$$\mu(m,k,i) = \zeta(m,k,i)\cdot\nu(m,k,i) \tag{51}$$

with:

$$\nu(m,k,i) = \frac{\sin\left(2\beta_2\pi^2 f_0^2(m-i)(k-i)N_sL_s\right)}{\sin\left(2\beta_2\pi^2 f_0^2(m-i)(k-i)L_s\right)}e^{j2\beta_2\pi^2 f_0^2(m-i)(k-i)(N_s-1)L_s} \tag{52}$$

Letting $f_0\to0$, we can then change the discrete-summation formula into integral form, whose result is shown in Sect. 3.

**Appendix B: Complete XCI formulas**

Here are the detailed expressions of the $\kappa_{mn}(f)$ contributions for XCI appearing in Eq. (18). The formulas for $\kappa_{11}(f)$ and $\kappa_{12}(f)$ were already shown in Sect. 4. The others are as follows:

$$\kappa_{21}(f) = \frac{32}{27} R_s^3 \int_{f_c-R_s/2}^{f_c+R_s/2} df_1 \int_{-R_s/2}^{+R_s/2} df_2\left|s_{\text{CUT}}(f_2)\right|^2\left|s_{\text{CUT}}(f_1+f_2-f)\right|^2\left|s_{\text{INT}}(f_1)\right|^2\left|\mu(f_1,f_2,f)\right|^2 \tag{53}$$

$$\kappa_{22}(f) = \frac{80}{81} R_s^2 \int_{f_c-R_s/2}^{f_c+R_s/2} df_1 \int_{-R_s/2}^{+R_s/2} df_2 \int_{-R_s/2}^{+R_s/2} df_2'\left|s_{\text{INT}}(f_1)\right|^2 s_{\text{CUT}}(f_2)s_{\text{CUT}}^*(f_2')$$
$$s_{\text{CUT}}^*(f_1+f_2-f)s_{\text{CUT}}(f_1+f_2'-f)\,\mu(f_1,f_2,f)\mu^*(f_1,f_2',f) \tag{54}$$

$$\kappa_{31}(f) = \frac{16}{27} R_s^3 \int\limits_{-R_s/2}^{+R_s/2} df_1 \int\limits_{-R_s/2}^{+R_s/2} df_2 \left| s_{\text{CUT}}(f_1) \right|^2 \left| s_{\text{CUT}}(f_2) \right|^2 \left| s_{\text{INT}}(f_1+f_2-f) \right|^2 \left| \mu(f_1,f_2,f) \right|^2 \quad (55)$$

$$\kappa_{32}(f) = \frac{16}{81} R_s^2 \int\limits_{-R_s/2}^{+R_s/2} df_1 \int\limits_{-R_s/2}^{+R_s/2} df_2 \int\limits_{-R_s/2}^{+R_s/2} df_2' \left| s_{\text{INT}}(f_1+f_2-f) \right|^2 s_{\text{CUT}}(f_1) s_{\text{CUT}}(f_2)$$
$$s_{\text{CUT}}^*(f_2') s_{\text{CUT}}^*(f_1+f_2-f_2') \mu(f_1,f_2,f) \mu^*(f_1+f_2-f_2',f_2',f)$$
$$(56)$$

$$\kappa_{41}(f) = \frac{16}{27} R_s^3 \int\limits_{f_c-R_s/2}^{f_c+R_s/2} df_1 \int\limits_{f_c-R_s/2}^{f_c+R_s/2} df_2 \left| s_{\text{INT}}(f_1) \right|^2 \left| s_{\text{INT}}(f_2) \right|^2 \left| s_{\text{INT}}(f_1+f_2-f) \right|^2 \left| \mu(f_1,f_2,f) \right|^2 \quad (57)$$

$$\kappa_{42}(f) = \frac{80}{81} R_s^2 \int\limits_{f_c-R_s/2}^{f_c+R_s/2} df_1 \int\limits_{f_c-R_s/2}^{f_c+R_s/2} df_2 \int\limits_{f_c-R_s/2}^{f_c+R_s/2} df_2' \left| s_{\text{INT}}(f_1) \right|^2 s_{\text{INT}}(f_2) s_{\text{INT}}^*(f_2')$$
$$s_{\text{INT}}^*(f_1+f_2-f) s_{\text{INT}}(f_1+f_2'-f) \mu(f_1,f_2,f) \mu^*(f_1,f_2',f)$$
$$+ \frac{16}{81} R_s^2 \int\limits_{f_c-R_s/2}^{f_c+R_s/2} df_1 \int\limits_{f_c-R_s/2}^{f_c+R_s/2} df_2 \int\limits_{f_c-R_s/2}^{f_c+R_s/2} df_2' \left| s_{\text{INT}}(f_1+f_2-f) \right|^2 s_{\text{INT}}(f_1) s_{\text{INT}}(f_2)$$
$$s_{\text{INT}}^*(f_1+f_2-f_2') s_{\text{INT}}^*(f_2') \mu(f_1,f_2,f) \mu^*(f_1+f_2-f_2',f_2',f)$$
$$(58)$$

$$\kappa_{43}(f) = \frac{16}{81} R_s \int\limits_{f_c-R_s/2}^{f_c+R_s/2} df_1 \int\limits_{f_c-R_s/2}^{f_c+R_s/2} df_2 \int\limits_{f_c-R_s/2}^{f_c+R_s/2} df_1' \int\limits_{f_c-R_s/2}^{f_c+R_s/2} df_2' s_{\text{INT}}(f_1) s_{\text{INT}}(f_2) s_{\text{INT}}^*(f_1+f_2-f)$$
$$s_{\text{INT}}^*(f_1') s_{\text{INT}}^*(f_2') s_{\text{INT}}(f_1'+f_2'-f) \mu(f_1,f_2,f) \mu^*(f_1',f_2',f)$$
$$(59)$$

## Appendix C: Derivation of XCI formulas

In this appendix, we present the derivation for the XCI formulas shown in Appendix B. In frequency domain, the signal model for two channels (dual polarization), i.e., the CUT and one INT channel, can be written as [7]:

$$\vec{E}(f) = E_x(f)\hat{x} + E_y(f)\hat{y} \quad (60)$$

where:

$$E_x(f) = \sqrt{f_0} \left( \sum_{n=-\infty}^{+\infty} v_{x,n} \delta(f-nf_0) + \sum_{n=-\infty}^{+\infty} \xi_{x,n} \delta(f-f_c-nf_0) \right) \quad (61)$$

$$E_y(f) = \sqrt{f_0} \left( \sum_{n=-\infty}^{+\infty} v_{y,n} \delta(f-nf_0) + \sum_{n=-\infty}^{+\infty} \xi_{y,n} \delta(f-f_c-nf_0) \right) \quad (62)$$

The random variables $v_{x,n}$, $v_{y,n}$, $\xi_{x,n}$ and $\xi_{y,n}$ are defined similar to [7], Appendix B, Eq. (36):

$$v_{x,n} = \sqrt{f_0} s_{\text{CUT},x}(nf_0) \sum_{\omega=0}^{W-1} a_{x,\omega} e^{-j\frac{2\pi}{W}\omega n} \quad , \quad v_{y,n} = \sqrt{f_0} s_{\text{CUT},y}(nf_0) \sum_{\omega=0}^{W-1} a_{y,\omega} e^{-j\frac{2\pi}{W}\omega n} \quad (63)$$

$$\xi_{x,n} = \sqrt{f_0} s_{\text{INT},x}(f_c+nf_0) \sum_{\omega=0}^{W-1} b_{x,\omega} e^{-j\frac{2\pi}{W}\omega n} \quad , \quad \xi_{y,n} = \sqrt{f_0} s_{\text{INT},y}(f_c+nf_0) \sum_{\omega=0}^{W-1} b_{y,\omega} e^{-j\frac{2\pi}{W}\omega n} \quad (64)$$

where $W$ is an integer number that can be chosen to be arbitrarily large.

Using the Manakov equation, the Kerr term at the fiber input on the $x$ polarization can be written as (see [10], Eqs. (28)-(29), (33)-(34) and (75)):

$$Q_{\text{NLI},x}(0,f) = -j\gamma \frac{8}{9} f_0^{3/2} \sum_{i=-\infty}^{+\infty} \delta(f - if_0)$$

$$\left[ \sum_{S_i} \left( v_{x,m} v_{x,n}^* v_{x,k} + v_{x,m} v_{y,n}^* v_{y,k} \right) \right.$$

$$+ \sum_{X1_i} \left( 2v_{x,m} \xi_{x,n}^* \xi_{x,k} + v_{y,m} \xi_{y,n}^* \xi_{x,k} + \xi_{y,m} \xi_{y,n}^* v_{x,k} \right)$$

$$+ \sum_{X2_i} \left( 2\xi_{x,m} v_{x,n}^* v_{x,k} + v_{y,m} v_{y,n}^* \xi_{x,k} + \xi_{y,m} v_{y,n}^* v_{x,k} \right)$$

$$+ \sum_{X3_i} \left( v_{x,m} \xi_{x,n}^* v_{x,k} + v_{y,m} \xi_{y,n}^* v_{x,k} \right) \tag{65}$$

$$+ \sum_{X4_i} \left( \xi_{x,m} \xi_{x,n}^* \xi_{x,k} + \xi_{y,m} \xi_{y,n}^* \xi_{x,k} \right)$$

$$\left. + \sum_{X5_i} \left( \xi_{x,m} v_{x,n}^* \xi_{x,k} + \xi_{y,m} v_{y,n}^* \xi_{x,k} \right) \right]$$

where:

$$X1_i \equiv S_i \equiv \left\{ (m,n,k) : (m-n+k)f_0 = if_0 \right\}$$
$$X2_i \equiv X4_i \equiv \left\{ (m,n,k) : (m-n+k)f_0 + f_c = if_0 \right\} \tag{66}$$
$$X3_i \equiv \left\{ (m,n,k) : (m-n+k)f_0 - f_c = if_0 \right\}$$
$$X5_i \equiv \left\{ (m,n,k) : (m-n+k)f_0 + 2f_c = if_0 \right\}$$

The first summation in the $Q_{\text{NLI},x}(0,f)$ formula is SCI, which is dealt with in Appendix A. The summation related to the index set X5 is always zero, as long as the channels do not spectrally overlap, i.e., as long as their separation is greater than $R_s$. We consider the spectral overlap case outside of the scope of this paper and therefore from now on we remove the summation related to X5.

Specifically, set X5 is generated by , $mf_0, kf_0 \in \text{INT}, nf_0 \in \text{CUT}$. It must be that $if_0 = mf_0 + kf_0 - nf_0 \in \text{CUT}$. In this paper, the bandwidth of each channel is equal to symbol rate $R_s$, and the central frequency of the INT is $f_c$, with ($f_c \geq R_s$). Therefore, we can write:

$$f_c - \frac{R_s}{2} \leq mf_0 \leq f_c + \frac{R_s}{2} \quad , \quad f_c - \frac{R_s}{2} \leq kf_0 \leq f_c + \frac{R_s}{2} \quad , \quad -\frac{R_s}{2} \leq nf_0 \leq +\frac{R_s}{2}$$

Combining these inequalities we get:

$$2f_c - \frac{3R_s}{2} \leq mf_0 + kf_0 - nf_0 \leq 2f_c + \frac{3R_s}{2}$$

Assuming $f_c > 0$, we can then remark that $2f_c - \frac{3R_s}{2} \geq \frac{R_s}{2}$, so that we can write:

$$\frac{R_s}{2} \leq mf_0 + kf_0 - nf_0 = if_0$$

where '=' holds only for $f_c = R_s$. Therefore, $if_0 \notin$ CUT and the contribution of the set X5 is zero. A similar conclusion can be found when assuming $f_c < 0$.

The resulting NLI field for the XCI component only, after a homogenous span of length $L_s$, can be written as:

$$E_{\text{XCI,x}}(L_s, f) = -j\frac{8}{9}f_0^{3/2}e^{-\alpha L_s}e^{-j2\beta_2\pi^2 i^2 f_0^2 L_s}\sum_{i=-\infty}^{+\infty}\delta(f - if_0)$$

$$\left[\sum_{\text{X1}_i}\zeta\left(k, m - \frac{f_c}{f_0}, n\right)\cdot\left(2v_{x,m}\xi_{x,n}^*\xi_{x,k} + v_{y,m}\xi_{y,n}^*\xi_{x,k} + \xi_{y,m}\xi_{y,n}^*v_{x,k}\right)\right.$$

$$+\sum_{\text{X2}_i}\zeta\left(k, m + \frac{f_c}{f_0}, n\right)\cdot\left(2\xi_{x,m}v_{x,n}^*v_{x,k} + v_{y,m}v_{y,n}^*\xi_{x,k} + \xi_{y,m}v_{y,n}^*v_{x,k}\right)$$

$$+\sum_{\text{X3}_i}\zeta\left(k - \frac{f_c}{f_0}, m - \frac{f_c}{f_0}, n\right)\cdot\left(v_{x,m}\xi_{x,n}^*v_{x,k} + v_{y,m}\xi_{y,n}^*v_{x,k}\right)$$

$$\left.+\sum_{\text{X4}_i}\zeta\left(k, m, n\right)\cdot\left(\xi_{x,m}\xi_{x,n}^*\xi_{x,k} + \xi_{y,m}\xi_{y,n}^*\xi_{x,k}\right)\right] \tag{67}$$

where $\zeta$ is defined as in Eq. (35):

$$\zeta\left(k, m, n\right) = \gamma\frac{1 - e^{-2\alpha L_s}e^{j4\pi^2\beta_2 f_0^2(k-n)(m-n)L_s}}{2\alpha - j4\pi^2\beta_2 f_0^2(k-n)(m-n)}$$

By 'homogeneous span' we mean that the fiber parameters are constant over the length $L_s$.

As for the field on the $y$ polarization, it can be found by swapping the subscripts $x$ and $y$. Therefore the total XCI PSD is:

$$G_{\text{XCI}}^{\text{EGN}}(f) = G_{\text{XCIX1}_i}(f) + G_{\text{XCIX2}_i}(f) + G_{\text{XCIX3}_i}(f) + G_{\text{XCIX4}_i}(f) \tag{68}$$

Since the only difference between these contributions is the cross-moments among six random variables, we just give the detailed derivation of the first contribution from set X1$_i$, which is related to the integration region X1 in Fig. 2.

In region X1, the XCI PSD is,

$$G_{\text{XCI,X1}_i}(f) = G_{\text{XCIX1}_{i,x}}(f) + G_{\text{XCIX1}_{i,y}}(f) \tag{69}$$

where:

$$G_{\text{XCIX1}_{i,x}}(f) = \frac{64}{81}f_0^3 e^{-2\alpha L}\sum_{i=-\infty}^{+\infty}\delta(f - if_0)\sum_{m,n,k\in\text{X1}_i}\sum_{m',n',k'\in\text{X1}_i}\zeta\left(k, m - \frac{f_c}{f_0}, n\right)\zeta^*\left(k', m' - \frac{f_c}{f_0}, n'\right)$$

$$\left[4\text{E}\left\{v_{x,m}v_{x,m'}^*\right\}\text{E}\left\{\xi_{x,n}^*\xi_{x,k}\xi_{x,n'}\xi_{x,k'}^*\right\} + 2\text{E}\left\{v_{x,m}v_{x,m'}^*\right\}\text{E}\left\{\xi_{x,n}^*\xi_{x,k}\right\}\text{E}\left\{\xi_{y,n'}\xi_{y,k'}^*\right\}\right.$$

$$+\text{E}\left\{v_{y,m}v_{y,m'}^*\right\}\text{E}\left\{\xi_{x,k}\xi_{x,k'}^*\right\}\text{E}\left\{\xi_{y,n}^*\xi_{y,n'}\right\} + \text{E}\left\{v_{x,m}v_{x,m'}^*\right\}\text{E}\left\{\xi_{y,n}^*\xi_{y,k}\xi_{y,n'}\xi_{y,k'}^*\right\}$$

$$\left.+2\text{E}\left\{v_{x,m}v_{x,m'}^*\right\}\text{E}\left\{\xi_{x,n}^*\xi_{x,k'}^*\right\}\text{E}\left\{\xi_{y,n}^*\xi_{y,k}\right\}\right] \tag{70}$$

We calculate these special expectations according to the formulas given in Appendix E, from which we can rewrite the XCI PSD as:

$$G_{\text{XCI.X}_l,x}(f) = \frac{64}{81} f_0^3 e^{-2\alpha L_s} \sum_{i=-\infty}^{+\infty} \delta(f - if_0) \sum_{m,n,k \in \text{X1}_i} \sum_{m',n',k' \in \text{X1}_i} \zeta\left(k, m - \frac{f_c}{f_0}, n\right) \zeta^*\left(k', m' - \frac{f_c}{f_0}, n'\right)$$

$$\left\{ 4R_s^3 \text{E}\left\{|a_x|^2\right\} \text{E}^2\left\{|b_x|^2\right\} \mathcal{P}_3 \delta_{m-m'} \delta_{n'-n} \delta_{k-k'} \right.$$

$$+ R_s^3 \text{E}\left\{|a_x|^2\right\} \text{E}^2\left\{|b_y|^2\right\} \mathcal{P}_4 \delta_{m-m'} \delta_{n'-n} \delta_{k-k'}$$

$$+ 4R_s^2 f_0 \text{E}\left\{|a_x|^2\right\} \text{E}^2\left\{|b_x|^2\right\} \left[\text{E}\left\{|b_x|^4\right\} / \text{E}^2\left\{|b_x|^2\right\} - 2\right] \mathcal{P}_3 \delta_{m-m'} \delta_{k-n-k'+n'} \tag{71}$$

$$+ R_s^2 f_0 \text{E}\left\{|a_x|^2\right\} \text{E}^2\left\{|b_y|^2\right\} \left[\text{E}\left\{|b_y|^4\right\} / \text{E}^2\left\{|b_y|^2\right\} - 2\right] \mathcal{P}_4 \delta_{m-m'} \delta_{k-n-k'+n'}$$

$$\left. + R_s^3 \text{E}\left\{|a_y|^2\right\} \text{E}\left\{|b_{x,0}|^2\right\} \text{E}\left\{|b_y|^2\right\} \mathcal{P}_5 \delta_{m-m'} \delta_{n'-n} \delta_{k-k'} \right\}$$

where:

$$\mathcal{P}_3 = s_{\text{CUT},x}(mf_0) s_{\text{INT},x}^*(f_c + nf_0) s_{\text{INT},x}(f_c + kf_0) s_{\text{CUT},x}^*(m'f_0) s_{\text{INT},x}(f_c + n'f_0) s_{\text{INT},x}^*(f_c + k'f_0) \tag{72}$$

$$\mathcal{P}_4 = s_{\text{CUT},x}(mf_0) s_{\text{INT},y}^*(f_c + nf_0) s_{\text{INT},y}(f_c + kf_0) s_{\text{CUT},x}^*(m'f_0) s_{\text{INT},y}(f_c + n'f_0) s_{\text{INT},y}^*(f_c + k'f_0) \tag{73}$$

$$\mathcal{P}_5 = s_{\text{CUT},y}(mf_0) s_{\text{INT},y}^*(f_c + nf_0) s_{\text{INT},x}(f_c + kf_0) s_{\text{CUT},y}^*(m'f_0) s_{\text{INT},y}(f_c + n'f_0) s_{\text{INT},x}^*(f_c + k'f_0) \tag{74}$$

In addition, we removed the terms with $k = n$ or $k' = n'$ because they can be shown to contribute a frequency-flat, constant phase shift which has no detrimental effect on transmission [10] (Eqs. (33)-(45)), [12], [13], [24].

If we also assume that:

$$\text{E}\left\{|a_x|^2\right\} = \text{E}\left\{|a_y|^2\right\} = \frac{1}{2}\text{E}\left\{|a|^2\right\} \quad , \quad \text{E}\left\{|b_x|^2\right\} = \text{E}\left\{|b_y|^2\right\} = \frac{1}{2}\text{E}\left\{|b|^2\right\} \tag{75}$$

$$s_{\text{CUT},x}(f) = s_{\text{CUT},y}(f) = s_{\text{CUT}}(f) \quad , \quad s_{\text{INT},x}(f) = s_{\text{INT},y}(f) = s_{\text{INT}}(f) \tag{76}$$

The simplified expression is,

$$G_{\text{XCI.X}_l,x}(f) = \frac{8}{81} f_0^3 e^{-2\alpha L_s} \text{E}\left\{|a|^2\right\} \text{E}^2\left\{|b|^2\right\} \sum_{i=-\infty}^{+\infty} \delta(f - if_0)$$

$$\sum_{m,n,k \in \text{X1}_i} \sum_{m',n',k' \in \text{X1}_i} \zeta\left(k, m - \frac{f_c}{f_0}, n\right) \zeta^*\left(k', m' - \frac{f_c}{f_0}, n'\right) \tag{77}$$

$$\left\{ 6R_s^3 \mathcal{P}_{\text{INT}} \delta_{m-m'} \delta_{k-k'} \delta_{n'-n} + 5R_s^2 f_0 \left[\text{E}\left\{|b|^4\right\} / \text{E}^2\left\{|b|^2\right\} - 2\right] \mathcal{P}_{\text{INT}} \delta_{m-m'} \delta_{k-n-k'+n'} \right\}$$

where:

$$\mathcal{P}_{\text{INT}} = s_{\text{CUT}}(mf_0) s_{\text{CUT}}^*(f_c + nf_0) s_{\text{INT}}(f_c + kf_0) s_{\text{CUT}}^*(m'f_0) s_{\text{INT}}(f_c + n'f_0) s_{\text{INT}}^*(f_c + k'f_0) \tag{78}$$

As for $G_{\text{XCI.X}_l,y}(f)$, it is identical to $G_{\text{XCI.X}_l,x}(f)$. Therefore, the XCI PSD in X1 is,

$$G_{\text{XCI.X}_l}(f) = \text{E}\left\{|a|^2\right\} \text{E}^2\left\{|b|^2\right\} \left[\chi_1(f) + \Phi_b \, \chi_2(f)\right] \tag{79}$$

where:

$$\chi_1(f) = \frac{16}{81} f_0^3 e^{-2\alpha L_s} \sum_{i=-\infty}^{+\infty} \delta(f-if_0) \sum_{m,n,k\in\mathrm{X1}_i} \sum_{m',n',k'\in\mathrm{X1}_i}$$

$$\zeta\left(k, m-\frac{f_c}{f_0}, n\right) \zeta^*\left(k', m'-\frac{f_c}{f_0}, n'\right) \cdot 6R_s^3 \mathcal{P}_{\mathrm{INT}} \delta_{m-m'} \delta_{k-k'} \delta_{n'-n}$$

$$= \frac{32}{27} f_0^3 e^{-2\alpha L_s} R_s^3 \sum_{i=-\infty}^{+\infty} \delta(f-if_0) \sum_m \sum_k |s_{\mathrm{CUT}}(mf_0)|^2 |s_{\mathrm{INT}}(f_c+kf_0)|^2$$

$$|s_{\mathrm{INT}}(f_c+[m+k-i]f_0)|^2 \left|\zeta\left(m,k+\frac{f_c}{f_0},i\right)\right|^2$$

(80)

$$\chi_2(f) = \frac{16}{81} f_0^3 e^{-2\alpha L_s} \sum_{i=-\infty}^{+\infty} \delta(f-if_0) \sum_{m,n,k\in\mathrm{X1}_i} \sum_{m',n',k'\in\mathrm{X1}_i}$$

$$\zeta\left(k, m-\frac{f_c}{f_0}, n\right) \zeta^*\left(k', m'-\frac{f_c}{f_0}, n'\right) \cdot 5R_s^3 f_0 \mathcal{P}_{\mathrm{INT}} \delta_{m-m'} \delta_{k-n-k'+n'}$$

$$= \frac{80}{81} f_0^4 e^{-2\alpha L_s} R_s^3 \sum_{i=-\infty}^{+\infty} \delta(f-if_0) \sum_m \sum_k \sum_{k'} |s_{\mathrm{CUT}}(mf_0)|^2 s_{\mathrm{INT}}(f_c+kf)$$

$$s_{\mathrm{INT}}^*(f_c+k'f_0) s_{\mathrm{INT}}^*(f_c+[m+k-i]f_0) s_{\mathrm{INT}}(f_c+[m+k'-i]f_0)$$

$$\zeta\left(m,k+\frac{f_c}{f_0},i\right) \zeta^*\left(m,k'+\frac{f_c}{f_0},i\right)$$

(81)

If identical spans using the same fiber type are assumed, with lumped amplifiers exactly compensating for the loss of each span, the XCI PSD is then:

$$G_{\mathrm{XCI}\mathrm{X1}_i}(f) = \mathrm{E}\left\{|a|^2\right\} \mathrm{E}^2\left\{|b|^2\right\} \left[\kappa_1(f) + \Phi_b \, \kappa_2(f)\right]$$

(82)

where:

$$\kappa_1(f) = \frac{32}{27} f_0^3 R_s^3 \sum_{i=-\infty}^{+\infty} \delta(f-if_0) \sum_m \sum_k |s_{\mathrm{CUT}}(mf_0)|^2 |s_{\mathrm{INT}}(f_c+kf_0)|^2$$

$$|s_{\mathrm{INT}}(f_c+[m+k-i]f_0)|^2 \left|\mu\left(m,k+\frac{f_c}{f_0},i\right)\right|^2$$

(83)

$$\kappa_2(f) = \frac{80}{81} f_0^4 R_s^3 \sum_{i=-\infty}^{+\infty} \delta(f-if_0) \sum_m \sum_k \sum_{k'} |s_{\mathrm{CUT}}(mf_0)|^2 s_{\mathrm{INT}}(f_c+kf_0)$$

$$s_{\mathrm{INT}}^*(f_c+k'f_0) s_{\mathrm{INT}}^*(f_c+[m+k-i]f_0) s_{\mathrm{INT}}(f_c+[m+k'-i]f_0)$$

$$\mu\left(m,k+\frac{f_c}{f_0},i\right) \mu^*\left(m,k'+\frac{f_c}{f_0},i\right)$$

(84)

Then transiting to integral format, we can get the final formulas shown in Sect. 4. As for the other contributions, they can be calculated through the same procedure, and related to different integration regions in Fig. 2. $G_{\mathrm{XCI}\mathrm{X2}_i}(f)$, $G_{\mathrm{XCI}\mathrm{X3}_i}(f)$ and $G_{\mathrm{XCI}\mathrm{X4}_i}(f)$ are induced by the integration regions X2, X3 and X4 respectively.

## Appendix D: Overview of MCI formulas derivation procedure

Figure 7 shows all the integration regions for a 9-channel system. The MCI regions are marked from M0 to M3. Increasing the number of channels does not create any new type of regions so, for this purpose, Fig. 7 can be considered an adequate generalization. The white-filled regions correspond to regions whose contribution is simply the GN-model; the other regions (M1-M3) have both a GN-model contribution and a correction term. Since all regions

have the GN-model contribution, we can generalize and say that MCI as a whole can be written as:

$$G_{\text{MCI}}^{\text{EGN}}(f) = G_{\text{MCI}}^{\text{GN}}(f) + G_{\text{MCI}}^{\text{corr}} \tag{85}$$

where $G_{\text{MCI}}^{\text{GN}}(f)$ is the MCI PSD according to the GN-model (present in M0-M4), and $G_{\text{MCI}}^{\text{corr}}$ is the correction found in the M1-M3 regions.

If all channels are assumed to have the same transmitted power, that is,

$$P_{\text{CUT}} = P_{\text{INT},i} = P_{\text{ch}} \quad i = -\left(N_{\text{ch}}-1\right)/2, \ldots, -1, 1, \ldots, \left(N_{\text{ch}}-1\right)/2 \tag{86}$$

where $N_{\text{ch}}$ (assumed odd) is the total number of channels and all INT channels are sitting symmetrically about CUT, then the MCI correction can be written as,

$$G_{\text{MCI}}^{\text{corr}} = \Phi_b P_{\text{ch}}^3 \left(\kappa_{\text{M1,2}}(f) + \kappa_{\text{M2,2}}(f) + \kappa_{\text{M1,2}}(f)\right) \tag{87}$$

The main difference between MCI formulas and their similar XCI formulas is the integration limits, therefore we need to find out the channels where the two triples $\left(f_1, f_2, f_3\right)$ and $\left(f_1', f_2', f_3'\right)$ are located.

- M1: similar to X1

Due to the symmetry, we evaluate MCI in the domains locating in the II quadrant, parallel to $f_2$. We can get:

$$f_1, f_1' \in \text{INT}_{-1}, \quad f_2, f_3, f_2', f_3' \in \text{INT}_n, n = 1, 2, \ldots, \left(N_{\text{ch}}-1\right)/2 \tag{88}$$

Therefore,

$$\kappa_{\text{M1,2}}(f) = 2 \cdot \frac{80}{81} R_s^2 \int_{-f_c - R_s/2}^{-f_c + R_s/2} df_1 \int_{nf_c - R_s/2}^{nf_c + R_s/2} df_2 \int_{nf_c - R_s/2}^{nf_c + R_s/2} df_2' \left| s_{\text{INT}_{-1}}(f_1) \right|^2$$
$$s_{\text{INT}_n}(f_2) s_{\text{INT}_n}^*(f_2') s_{\text{INT}_n}^*(f_1 + f_2 - f) s_{\text{INT}_n}(f_1 + f_2' - f) \mu\left(f_1, f_2, f\right) \mu^*\left(f_1, f_2', f\right) \tag{89}$$
$$\left(\text{with } n = 1, 2, \ldots, \left(N_{\text{ch}}-1\right)/2\right)$$

- M2: similar to X1

For the domains locating in the I quadrant, parallel to $f_2$. We can get:

$$f_1, f_1' \in \text{INT}_1, \quad f_2, f_3, f_2', f_3' \in \text{INT}_n, n = 2, 3, \ldots, \left(N_{\text{ch}}-1\right)/2 \tag{90}$$

Therefore,

$$\kappa_{\text{M2,2}}(f) = 2 \cdot \frac{80}{81} R_s^2 \int_{f_c - R_s/2}^{f_c + R_s/2} df_1 \int_{nf_c - R_s/2}^{nf_c + R_s/2} df_2 \int_{nf_c - R_s/2}^{nf_c + R_s/2} df_2' \left| s_{\text{INT}_1}(f_1) \right|^2$$
$$s_{\text{INT}_n}(f_2) s_{\text{INT}_n}^*(f_2') s_{\text{INT}_n}^*(f_1 + f_2 - f) s_{\text{INT}_n}(f_1 + f_2' - f) \mu\left(f_1, f_2, f\right) \mu^*\left(f_1, f_2', f\right) \tag{91}$$
$$\left(\text{with } n = 2, 3, \ldots, \left(N_{\text{ch}}-1\right)/2\right)$$

- M3: similar to X3

For the domains locating in the I quadrant, we can get:

$$f_3, f_3' \in \text{INT}_n, \ n = 2, 3, \ldots, (N_{ch} - 1)/2$$

$$f_1, f_2, f_1', f_2' \in \text{INT}_m, \ m = \begin{cases} n/2, & n \text{ is even} \\ (n \pm 1)/2, & n \text{ is odd} \end{cases} \tag{92}$$

Therefore,

$$\kappa_{M3,2}(f) = 2 \cdot \frac{16}{81} R_s^2 \int_{mf_c - R_s/2}^{mf_c + R_s/2} df_1 \int_{mf_c - R_s/2}^{mf_c + R_s/2} df_2 \int_{mf_c - R_s/2}^{mf_c + R_s/2} df_2' \left| s_{\text{INT}_n}(f_1 + f_2 - f) \right|^2$$

$$s_{\text{INT}_m}(f_1) s_{\text{INT}_m}^*(f_2) s_{\text{INT}_m}^*(f_2') s_{\text{INT}_m}^*(f_1 + f_2 - f_2') \mu(f_1, f_2, f) \mu^*(f_1 + f_2 - f_2', f_2', f) \tag{93}$$

$$\left( \text{with} \begin{cases} n = 2, 3, \ldots, (N_{ch} - 1)/2 \\ m = \begin{cases} n/2, & n \text{ is even} \\ (n \pm 1)/2, & n \text{ is odd} \end{cases} \end{cases} \right)$$

**Appendix E: Some relevant $\xi_n$ moments calculation**

In frequency domain, the transmitted symbol sequence in a generic interfering channel can be written as:

$$\xi_n = \sqrt{f_0} s(n f_0) \sum_{\omega=0}^{W-1} b_\omega e^{-j \frac{2\pi}{W} \omega n} \tag{94}$$

Its 2$^\text{nd}$-order moment is well known as,

$$\text{E}\left\{ \xi_m \xi_n^* \right\} = R_s \left| s(m f_0) \right|^2 \text{E}\left\{ |b|^2 \right\} \left( \delta_{m-n+pW} \right)_{p=0,\pm 1,\ldots} \tag{95}$$

Its 4$^\text{th}$-order moment is,

$$\text{E}\left\{ \xi_m \xi_n^* \xi_{m'}^* \xi_{n'} \right\}$$

$$= \text{E}\left\{ \sqrt{f_0} s(m f_0) \sum_{\omega_1=0}^{W-1} b_{\omega_1} e^{-j \frac{2\pi}{W} \omega_1 m} \cdot \sqrt{f_0} s^*(n f_0) \sum_{\omega_2=0}^{W-1} b_{\omega_2}^* e^{j \frac{2\pi}{W} \omega_2 n} \right.$$

$$\left. \cdot \sqrt{f_0} s^*(m' f_0) \sum_{\omega_3=0}^{W-1} b_{\omega_3}^* e^{j \frac{2\pi}{W} \omega_3 m'} \cdot \sqrt{f_0} s(n' f_0) \sum_{\omega_4=0}^{W-1} b_{\omega_4} e^{-j \frac{2\pi}{W} \omega_4 n'} \right\} \tag{96}$$

$$= f_0^2 \mathcal{P}_{mnm'n'} \sum_{\omega_1=0}^{W-1} \sum_{\omega_2=0}^{W-1} \sum_{\omega_3=0}^{W-1} \sum_{\omega_4=0}^{W-1} \text{E}\left\{ b_{\omega_1} b_{\omega_2}^* b_{\omega_3}^* b_{\omega_4} \right\} e^{-j \frac{2\pi}{W} (\omega_1 m - \omega_2 n - \omega_3 m' + \omega_4 n')}$$

Where $\mathcal{P}_{mnm'n'} = s(m f_0) s^*(n f_0) s^*(m' f_0) s(n' f_0)$.

The calculation of the 4$^\text{th}$-order correlation of RV $b$ can be split to two groups:

- $\omega_1 = \omega_2 = \omega_3 = \omega_4$: the four-summation can be reduced to one single-summation.

$$\text{E}^{(i)}\left\{ \xi_n \xi_n^* \xi_m^* \xi_{n'} \right\} = f_0^2 \mathcal{P}_{mnm'n'} \text{E}\left\{ |b|^4 \right\} \sum_{\omega_1=0}^{W-1} e^{-j \frac{2\pi}{W} \omega_1 (m-n-m'+n')}$$

$$= R_s f_0 \text{E}\left\{ |b|^4 \right\} \mathcal{P}_{mnm'n'} \left( \delta_{m-n-m'+n'+pW} \right)_{p=0,\pm 1,\ldots} \tag{97}$$

where we used the stationary $\mathrm{E}\left\{|b|^4\right\} = \mathrm{E}\left\{\left|b_{\omega_1}\right|^4\right\}$.

- $\left\{\omega_i = \omega_j, i = 1,4, j = 2,3\right\}, \omega_1 \neq \omega_4$ : the four-summation can be reduced to two double-summation.

(2.1) $\left\{\omega_1 = \omega_2, \omega_4 = \omega_3, \omega_1 \neq \omega_4\right\}$

$$\mathrm{E}^{(2.1)}\left\{\xi_m \xi_n^* \xi_{m'}^* \xi_{n'}\right\}$$

$$= f_0^2 \mathcal{P}_{mnm'n'} \cdot \mathrm{E}^2\left\{|b|^2\right\} \sum_{\omega_1=0}^{W-1} e^{-j\frac{2\pi}{W}\omega_1(m-n)} \sum_{\omega_4=0, \ \omega_4 \neq \omega_1}^{W-1} e^{-j\frac{2\pi}{W}\omega_3(-m'+n')}$$

$$= f_0^2 \mathrm{E}^2\left\{|b|^2\right\} \mathcal{P}_{mnm'n'}\left(W^2 \delta_{m-n+pW}\delta_{-m'+n'+pW} - W\delta_{m-n-m'+n'+pW}\right)_{p=0,\pm1,\ldots} \quad (98)$$

$$= R_s^2 \mathrm{E}^2\left\{|b|^2\right\} \mathcal{P}_{mnm'n'}\left(\delta_{m-n+pW}\delta_{-m'+n'+pW}\right)_{p=0,\pm1,\ldots}$$

$$- R_s f_0 \mathrm{E}^2\left\{|b|^2\right\} \mathcal{P}_{mnm'n'}\left(\delta_{m-n-m'+n'+pW}\right)_{p=0,\pm1,\ldots}$$

(2.2) $\left\{\omega_1 = \omega_3, \omega_4 = \omega_2, \omega_1 \neq \omega_4\right\}$

$$\mathrm{E}^{(2.2)}\left\{\xi_m \xi_n^* \xi_{m'}^* \xi_{n'}\right\}$$

$$= R_s^2 \mathrm{E}^2\left\{|b|^2\right\} \mathcal{P}_{mnm'n'}\left(\delta_{m-m'+pW}\delta_{-n+n'+pW}\right)_{p=0,\pm1,\ldots} \quad (99)$$

$$- R_s f_0 \mathrm{E}^2\left\{|b|^2\right\} \mathcal{P}_{mnm'n'}\left(\delta_{m-n-m'+n'+pW}\right)_{p=0,\pm1,\ldots}$$

Putting these contributions together, we can get,

$$\mathrm{E}\left\{\xi_m \xi_n^* \xi_{m'}^* \xi_{n'}\right\} = \mathrm{E}^{(1)}\left\{\xi_m \xi_n^* \xi_{m'}^* \xi_{n'}\right\} + \mathrm{E}^{(2)}\left\{\xi_m \xi_n^* \xi_{m'}^* \xi_{n'}\right\} + \mathrm{E}^{(3)}\left\{\xi_m \xi_n^* \xi_{m'}^* \xi_{n'}\right\}$$

$$= R_s^2 \mathrm{E}^2\left\{|b|^2\right\} \mathcal{P}_{mnm'n'}\left(\delta_{m-n+pW}\delta_{n'-m'+pW} + \delta_{m-m'+pW}\delta_{n'-n+pW}\right)_{p=0,\pm1,\ldots} \quad (100)$$

$$+ R_s f_0 \left[\mathrm{E}\left\{|b|^4\right\} - 2\mathrm{E}^2\left\{|b|^2\right\}\right] \mathcal{P}_{mnm'n'}\left(\delta_{m-n-m'+n'+pW}\right)_{p=0,\pm1,\ldots}$$

Its 6th-order moment is:

$$\mathrm{E}\left\{\xi_m \xi_n^* \xi_k \xi_{m'}^* \xi_{n'} \xi_{k'}^*\right\}$$

$$= f_0^3 \mathcal{P}_{mnkm'n'k'} \sum_{\omega_1=0}^{W-1} \sum_{\omega_2=0}^{W-1} \sum_{\omega_3=0}^{W-1} \sum_{\omega_4=0}^{W-1} \sum_{\omega_5=0}^{W-1} \sum_{\omega_6=0}^{W-1} \quad (101)$$

$$\mathrm{E}\left\{b_{\omega_1} b_{\omega_2}^* b_{\omega_3} b_{\omega_4}^* b_{\omega_5} b_{\omega_6}^*\right\} e^{-j\frac{2\pi}{W}(\omega_1 m - \omega_2 n + \omega_3 k - \omega_4 m' + \omega_5 n' - \omega_6 k')}$$

Where $\mathcal{P}_{mnkm'n'k'} = s(mf_0) s^*(nf_0) s(kf_0) s^*(m'f_0) s(n'f_0) s^*(k'f_0)$, and the 6th-order correlation of RV $b$ can be split to three groups.

- $\omega_1 = \omega_2 = \omega_3 = \omega_4 = \omega_5 = \omega_6$ : the six-summation can be reduced to one single-summation.

$$\mathrm{E}^{(1)}\left\{\xi_m \xi_n^* \xi_k \xi_{m'}^* \xi_{n'} \xi_{k'}^*\right\}$$

$$= f_0^3 \boldsymbol{\mathcal{P}}_{mnkm'n'k'} \mathrm{E}\left\{|b|^6\right\} \sum_{\omega_1=0}^{W-1} e^{-j\frac{2\pi}{W}\omega_1\left(m-n+k-m'+n'-k'\right)} \quad (102)$$

$$= R_s f_0^2 \mathrm{E}\left\{|a_0|^6\right\} \boldsymbol{\mathcal{P}}_{mnkm'n'k'} \left(\delta_{m-n+k-m'+n'-k'+pW}\right)_{p=0,\pm1,\ldots}$$

- Two of them are identical, and the other four are identical, thus the six-summation can be reduced to nine dual-summation.

  (2.1) $\omega_1=\omega_2$, $\omega_3=\omega_4=\omega_5=\omega_6$, $\omega_1\neq\omega_3$

  (2.2) $\omega_1=\omega_4$, $\omega_3=\omega_2=\omega_5=\omega_6$, $\omega_1\neq\omega_3$

  (2.3) $\omega_1=\omega_6$, $\omega_3=\omega_4=\omega_5=\omega_2$, $\omega_1\neq\omega_3$

  (2.4) $\omega_3=\omega_2$, $\omega_1=\omega_4=\omega_5=\omega_6$, $\omega_3\neq\omega_1$

  (2.5) $\omega_3=\omega_4$, $\omega_1=\omega_2=\omega_5=\omega_6$, $\omega_3\neq\omega_1$

  (2.6) $\omega_3=\omega_6$, $\omega_1=\omega_4=\omega_5=\omega_2$, $\omega_3\neq\omega_1$

  (2.7) $\omega_5=\omega_2$, $\omega_1=\omega_4=\omega_3=\omega_6$, $\omega_5\neq\omega_1$

  (2.8) $\omega_5=\omega_4$, $\omega_1=\omega_2=\omega_3=\omega_6$, $\omega_5\neq\omega_1$

  (2.9) $\omega_5=\omega_6$, $\omega_1=\omega_4=\omega_3=\omega_2$, $\omega_5\neq\omega_1$

Here we only give the procedure for calculating (2.1).

$$\mathrm{E}^{(2.1)}\left\{\xi_m \xi_n^* \xi_k \xi_{m'}^* \xi_{n'} \xi_{k'}^*\right\}$$

$$= f_0^3 \boldsymbol{\mathcal{P}}_{mnkm'n'k'} \mathrm{E}\left\{|b|^2\right\} \sum_{\omega_1=0}^{W-1} e^{-j\frac{2\pi}{W}\omega_1(m-n)} \mathrm{E}\left\{|b|^4\right\} \sum_{\omega_3=0,\,\omega_3\neq\omega_1}^{W-1} e^{-j\frac{2\pi}{W}\omega_3(k-m'+n'-k')}$$

$$= f_0^3 \mathrm{E}\left\{|b|^2\right\} \mathrm{E}\left\{|b|^4\right\} \boldsymbol{\mathcal{P}}_{mnkm'n'k'} \left(W^2 \delta_{m-n+pW}\delta_{k-m'+n'-k'+pW} - W\delta_{m-n+k-m'+n'-k'+pW}\right)_{p=0,\pm1,\ldots}$$

$$= R_s^2 \mathrm{E}\left\{|b|^2\right\} \mathrm{E}\left\{|b|^4\right\} \boldsymbol{\mathcal{P}}_{mnkm'n'k'} \left(\delta_{m-n+pW}\delta_{k-m'+n'-k'+pW}\right)_{p=0,\pm1,\ldots}$$

$$- R_s f_0^2 \mathrm{E}\left\{|b|^2\right\} \mathrm{E}\left\{|b|^4\right\} \boldsymbol{\mathcal{P}}_{mnkm'n'k'} \left(\delta_{m-n+k-m'+n'-k'+pW}\right)_{p=0,\pm1,\ldots}$$

$$(103)$$

- $\left\{\omega_i=\omega_j, i=1,3,5,\ j=2,4,6\right\}$, $\omega_1\neq\omega_3\neq\omega_5$: the six-summation can be reduced to six triple-summation.

  (3.1) $\omega_1=\omega_2$, $\omega_3=\omega_4$, $\omega_5=\omega_6$     (3.2) $\omega_1=\omega_2$, $\omega_3=\omega_6$, $\omega_5=\omega_4$

  (3.3) $\omega_1=\omega_4$, $\omega_3=\omega_2$, $\omega_5=\omega_6$     (3.4) $\omega_1=\omega_4$, $\omega_3=\omega_6$, $\omega_5=\omega_2$

  (3.5) $\omega_1=\omega_6$, $\omega_3=\omega_2$, $\omega_5=\omega_4$     (3.6) $\omega_1=\omega_6$, $\omega_3=\omega_4$, $\omega_5=\omega_2$

Here we only give the procedure for calculating (3.1).

$$
\mathrm{E}^{(3.1)}\left\{\xi_m\xi_n^*\xi_k\xi_m^*\xi_n\xi_{k'}^*\right\}
$$

$$
= f_0^3\,\boldsymbol{\mathcal{P}}_{mnkm'n'k'}\mathrm{E}^3\left\{|b|^2\right\}\sum_{\omega_1=0}^{W-1}e^{-j\frac{2\pi}{W}\omega_1(m-n)}\sum_{\omega_2=0}^{W-1}e^{-j\frac{2\pi}{W}\omega_2(k-m')}\sum_{\omega_3=0}^{W-1}e^{-j\frac{2\pi}{W}\omega_3(n'-k')}
$$

$$
= f_0^3\,\boldsymbol{\mathcal{P}}_{mnkm'n'k'}\mathrm{E}^3\left\{|b|^2\right\}\Big(W^3\delta_{m-n+pW}\delta_{k-m'+pW}\delta_{n'-k'+pW}-W^2\delta_{m-n+pW}\delta_{k-m'+n'-k'+pW}-
$$

$$
-W^2\delta_{k-m'+pW}\delta_{m-n+n'-k'+pW}-W^2\delta_{n'-k'+pW}\delta_{m-n+k-m'+pW}+2W\delta_{m-n+k-m'+n'-k'+pW}\Big)_{p=0,\pm1,\dots}
$$

$$
= R_s^3\,\boldsymbol{\mathcal{P}}_{mnkm'n'k'}\mathrm{E}^3\left\{|b|^2\right\}\Big(\delta_{m-n+pW}\delta_{k-m'+pW}\delta_{n'-k'+pW}\Big)_{p=0,\pm1,\dots}-R_s^2 f_0\boldsymbol{\mathcal{P}}_{mnkm'n'k'}\mathrm{E}^3\left\{|b|^2\right\}
$$

$$
\cdot\Big(\delta_{m-n+pW}\delta_{k-m'+n'-k'+pW}+\delta_{k-m'+pW}\delta_{m-n+n'-k'+pW}+\delta_{n'-k'+pW}\delta_{m-n+k-m'+pW}\Big)_{p=0,\pm1,\dots}
$$

$$
+2R_s f_0^2\,\boldsymbol{\mathcal{P}}_{mnkm'n'k'}\mathrm{E}^3\left\{|b|^2\right\}\Big(\delta_{m-n+k-m'+n'-k'+pW}\Big)_{p=0,\pm1,\dots}
$$

$$
(104)
$$

Putting all contributions together, we can get,

$$
\mathrm{E}\left\{\xi_m\xi_n^*\xi_k\xi_m^*\xi_n\xi_{k'}^*\right\}
$$

$$
= R_s f_0^2\mathrm{E}^3\left\{|b|^2\right\}\left[\mathrm{E}\left\{|b|^6\right\}\Big/\mathrm{E}^3\left\{|b|^2\right\}-9\,\mathrm{E}\left\{|b|^4\right\}\Big/\mathrm{E}^2\left\{|b|^2\right\}+12\right]
$$

$$
\cdot\boldsymbol{\mathcal{P}}_{mnkm'n'k'}\left(\delta_{m-n+k-m'+n'-k'+pW}\right)_{p=0,\pm1,\dots}+R_s^2 f_0\mathrm{E}^3\left\{|b|^2\right\}
$$

$$
\cdot\left[\mathrm{E}\left\{|b|^4\right\}\Big/\mathrm{E}^2\left\{|b|^2\right\}-2\right]\boldsymbol{\mathcal{P}}_{mnkm'n'k'}\left(\delta_{m-n+pW}\delta_{k-m'+n'-k'+pW}\right.
$$

$$
+\delta_{m-m'+pW}\delta_{k-n+n'-k'+pW}+\delta_{m-k'+pW}\delta_{k-n-m'+n'+pW}
$$

$$
+\delta_{k-n+pW}\delta_{m-m'+n'-k'+pW}+\delta_{k-m'+pW}\delta_{m-n+n'-k'+pW}
$$

$$
+\delta_{k-k'+pW}\delta_{m-n-m'+n'+pW}+\delta_{n'-n+pW}\delta_{m+k-m'-k'+pW}
$$

$$
+\delta_{n'-m'+pW}\delta_{m-n+k-k'+pW}+\delta_{n'-k'+pW}\delta_{m-n+k-m'+pW}\Big)_{p=0,\pm1,\dots}
$$

$$
+R_s^3\mathrm{E}^3\left\{|b|^2\right\}\boldsymbol{\mathcal{P}}_{mnkm'n'k'}\left(\delta_{m-n+pW}\delta_{k-m'+pW}\delta_{n'-k'+pW}\right.
$$

$$
+\delta_{m-n+pW}\delta_{k-k'+pW}\delta_{n'-m'+pW}+\delta_{m-m'+pW}\delta_{k-n+pW}\delta_{n'-k'+pW}
$$

$$
+\delta_{m-m'+pW}\delta_{k-k'+pW}\delta_{n'-n+pW}+\delta_{m-k'+pW}\delta_{k-n+pW}\delta_{n'-m'+pW}
$$

$$
+\delta_{m-k'+pW}\delta_{k-m'+pW}\delta_{n'-n+pW}\Big)_{p=0,\pm1,\dots}
$$

$$
(105)
$$

If we choose ideally rectangular spectrum, the parameter $p$ is equal to 0. In this paper we assume that the channels are rectangular or almost rectangular and neglect the contribution of the terms arising when $p\neq0$, similar to what was done in [24]. Investigating the impact of this approximation, when channel spectra are significantly far from rectangular, is left for future investigation.

## Appendix F: Analytical complexity of the EGN model terms

As shown in Eq. (1), the EGN-model consist of a GN-model term $G_{\mathrm{NLI}}^{\mathrm{GN}}(f)$ and a 'correction term' $G_{\mathrm{NLI}}^{\mathrm{corr}}(f)$. All the contributions making up the GN-model term consist of double integrals over $f_1, f_2$. The contributions of the EGN-model correction term $G_{\text{NLI}}^{\text{corr}}(f)$ are instead expressed as either triple or quadruple integrals. This seems to suggest that the numerical integration of the $G_{\text{NLI}}^{\text{corr}}(f)$ contributions may be quite challenging.

In reality, the $G_{\text{NLI}}^{\text{corr}}(f)$ contributions can be shown to always require only a double integral to be evaluated. For instance, one of the correction terms for SCI which has a triple integral is Eq. (8):

$$
\begin{aligned}
\kappa_2(f) = \frac{80}{81} R_s^2 \int_{-R_s/2}^{+R_s/2} df_1 \int_{-R_s/2}^{+R_s/2} df_2 \int_{-R_s/2}^{+R_s/2} df_2' \\
\left| s_{\text{CUT}}(f_1) \right|^2 s_{\text{CUT}}(f_2) s_{\text{CUT}}^*(f_2') s_{\text{CUT}}^*(f_1+f_2-f) s_{\text{CUT}}(f_1+f_2'-f) \mu(f_1,f_2,f) \mu^*(f_1,f_2',f) \\
+ \frac{16}{81} R_s^2 \int_{-R_s/2}^{+R_s/2} df_1 \int_{-R_s/2}^{+R_s/2} df_2 \int_{-R_s/2}^{+R_s/2} df_2' \\
\left| s_{\text{CUT}}(f_1+f_2-f) \right|^2 s_{\text{CUT}}(f_1) s_{\text{CUT}}(f_2) s_{\text{CUT}}^*(f_1+f_2-f_2') s_{\text{CUT}}^*(f_2') \mu(f_1,f_2,f) \mu^*(f_1+f_2-f_2',f_2',f)
\end{aligned}
\tag{106}
$$

The term preceded by $80/81$ in the equation above, which we will call $\kappa_{21}(f)$, can be re-written as:

$$
\begin{aligned}
\kappa_{21}(f) = \frac{80}{81} R_s^2 \int_{-R_s/2}^{+R_s/2} df_1 \left| s_{\text{CUT}}(f_1) \right|^2 \\
\int_{-R_s/2}^{+R_s/2} df_2 \, s_{\text{CUT}}(f_2) s_{\text{CUT}}^*(f_1+f_2-f) \mu(f_1,f_2,f) \\
\int_{-R_s/2}^{+R_s/2} df_2' \, s_{\text{CUT}}^*(f_2') s_{\text{CUT}}(f_1+f_2'-f) \mu^*(f_1,f_2',f) \\
= \frac{80}{81} R_s^2 \int_{-R_s/2}^{+R_s/2} df_1 \left| s_{\text{CUT}}(f_1) \right|^2 \cdot \left| \int_{-R_s/2}^{+R_s/2} df_2 \, s_{\text{CUT}}(f_2) s_{\text{CUT}}^*(f_1+f_2-f) \mu(f_1,f_2,f) \right|^2
\end{aligned}
\tag{107}
$$

In other words, the second and third integral are the same integral, except for a complex conjugation, so that only one integration is needed to obtain both.

For the term preceded by $16/81$, which we will call $\kappa_{22}(f)$, we replace the integration variable $f_1$ with $f_3 = f_1+f_2-f$, that is $f_1 = f_3-f_2+f$. Then:

$$
\begin{aligned}
\kappa_{22}(f) = \frac{16}{81} R_s^2 \int_{-R_s/2}^{+R_s/2} df_3 \int_{-R_s/2}^{+R_s/2} df_2 \int_{-R_s/2}^{+R_s/2} df_2' \left| s_{\text{CUT}}(f_3) \right|^2 s_{\text{CUT}}(f_3-f_2+f) \\
s_{\text{CUT}}(f_2) s_{\text{CUT}}^*(f_3-f_2'+f) s_{\text{CUT}}^*(f_2') \mu(f_3-f_2+f,f_2,f) \mu^*(f_3-f_2'+f,f_2',f) \\
= \frac{16}{81} R_s^2 \int_{-R_s/2}^{+R_s/2} df_3 \left| s_{\text{CUT}}(f_3) \right|^2 \\
\int_{-R_s/2}^{+R_s/2} df_2 \, s_{\text{CUT}}(f_2) s_{\text{CUT}}(f_3-f_2+f) \mu(f_3-f_2+f,f_2,f) \\
\int_{-R_s/2}^{+R_s/2} df_2' \, s_{\text{CUT}}^*(f_2') s_{\text{CUT}}^*(f_3-f_2'+f) \mu^*(f_3-f_2'+f,f_2',f) \\
= \frac{16}{81} R_s^2 \int_{-R_s/2}^{+R_s/2} df_3 \left| s_{\text{CUT}}(f_3) \right|^2 \cdot \left| \int_{-R_s/2}^{+R_s/2} df_2 \, s_{\text{CUT}}(f_2) s_{\text{CUT}}(f_3-f_2+f) \mu(f_3-f_2+f,f_2,f) \right|^2
\end{aligned}
\tag{108}
$$

Again, the second and third integrals are the same integral, except for a complex conjugation, so that only one integration is actually needed to obtain both.

One of the correction terms for SCI has a quadruple integral Eq. (9):

$$
\begin{aligned}
\kappa_3(f) = \frac{16}{81} R_s &\int\limits_{-R_s/2}^{+R_s/2} df_1 \int\limits_{-R_s/2}^{+R_s/2} df_2 \int\limits_{-R_s/2}^{+R_s/2} df_1' \int\limits_{-R_s/2}^{+R_s/2} df_2' s_{\text{CUT}}(f_1) s_{\text{CUT}}(f_2) \\
&s_{\text{CUT}}^*(f_1+f_2-f) s_{\text{CUT}}^*(f_1') s_{\text{CUT}}^*(f_2') s_{\text{CUT}}(f_1'+f_2'-f) \mu(f_1,f_2,f) \mu^*(f_1',f_2',f) \\
= \frac{16}{81} R_s &\int\limits_{-R_s/2}^{+R_s/2} df_1 \int\limits_{-R_s/2}^{+R_s/2} df_2 s_{\text{CUT}}(f_1) s_{\text{CUT}}(f_2) s_{\text{CUT}}^*(f_1+f_2-f) \mu(f_1,f_2,f) \\
&\int\limits_{-R_s/2}^{+R_s/2} df_1' \int\limits_{-R_s/2}^{+R_s/2} df_2' s_{\text{CUT}}^*(f_1') s_{\text{CUT}}^*(f_2') s_{\text{CUT}}(f_1'+f_2'-f) \mu^*(f_1',f_2',f) \\
= \frac{16}{81} R_s &\left| \int\limits_{-R_s/2}^{+R_s/2} df_1 \int\limits_{-R_s/2}^{+R_s/2} df_2 s_{\text{CUT}}(f_1) s_{\text{CUT}}(f_2) s_{\text{CUT}}^*(f_1+f_2-f) \mu(f_1,f_2,f) \right|^2
\end{aligned} \tag{109}
$$

Here, it turns out that the first two integrals together are the complex conjugate of the third and fourth, so that a double integration only is needed to assess the whole contribution.

Similar manipulations can be used to show that all other EGN-model contributions, including XCI and MCI have an inherent complexity that is just that of a double-integral.

This property is clearly important and we exploited it in the numerical evaluation software that we used. It is also possible that more analytical manipulation can be carried out to further reduce the integration complexity. For instance in [11] a hyperbolic variable substitution proved quite effective for the GN-model. However, we have not yet carried out a similar investigation for the EGN model and leave this topic for possible future research.